\documentclass[namedreferences,hyperref,optionalrh,solaromanenum]{spr-sola}

\usepackage{graphicx}                    % For eps figures, newer & more powerfull
\usepackage{color}                       % For color text: \color command
\chardef\us=`\_

\begin{document}
\begin{frontmatter}

%\title{The Distribution of Photospheric Magnetic Flux Features\\
\title{Solar Cycle Variation of the Distribution of Photospheric Magnetic Flux Features}

%%%%%%%%%%%%%%%%%%%%%%%%%%%%%%%%%%%%%%%%%%%%%%%%%%%
%% Authors Names
%
% \author[addressref={},corref,email={}]{\inits{}\fnm{}\lnm{}\orcid{}}
\author[addressref={aff1},email={cnn@st-andrews.ac.uk}]{\inits{C.N.}\fnm{Callan N. }\lnm{Noble}
\orcid{0000-0003-4694-8537}}
\author[addressref=aff1,email={cep@st-andrews.ac.uk}]{\inits{C.E.}\fnm{Clare E.}~\lnm{Parnell}
\orcid{0000-0002-5694-9069}}
\author[addressref=aff1,corref,email={tn3@st-andrews.ac.uk}]{\inits{T.}\fnm{Thomas}~\lnm{Neukirch}
\orcid{0000-0002-7597-4980}}

%%%%%%%%%%%%%%%%%%%%%%%%%%%%%%%%%%%%%%%%%%%%%%%%%%%
%% Runningheads
%
\runningauthor{C.N. Noble \textit{et al.}}

\runningtitle{The Distribution of Photospheric Magnetic Flux Features}

%%%%%%%%%%%%%%%%%%%%%%%%%%%%%%%%%%%%%%%%%%%%%%%%%%%
%% Affilations 
%% id should be the same with \author addressref value.
\address[id=aff1]{School of Mathematics and Statistics, University of St Andrews, North Haugh, St Andrews, KY16 9SS, UK}

%%%%%%%%%%%%%%%%%%%%%%%%%%%%%%%%%%%%%%%%%%%%%%%%%%%
%%% Abstract 
\begin{abstract}

We use statistical tools to analyse data from the Solar Dynamics Observatory
Helioseismic and Magnetic Imager to determine the distribution of the magnetic flux of photospheric 
magnetic features and its variation over a full solar cycle. 
In particular, we use statistical figures of merit to test 
how well different types of probability 
density
function represent the 
magnetic flux distribution inferred from the data and how their shape changes over the solar cycle.
Our analysis shows that a double power law provides the best representation of the data over the full solar cycle and
we present the dependence of the power law exponents on the phase of the solar cycle.
The nature of the observed flux distributions at different times during the 
solar cycle is significant because it could be used to try and infer 
information about solar magnetic field generation mechanisms. We discuss potential implications of a double power law distribution for solar magnetic field generation.

\end{abstract}

%%%%%%%%%%%%%%%%%%%%%%%%%%%%%%%%%%%%%%%%%%%%%%%%%%%
%% Keywords
%
\keywords{Magnetic fields, Photosphere; Solar Cycle, Observations; Sunspots, Statistics}

\end{frontmatter}

%\newpage
%-------------------------------------------------

%%%%%%%%%%%%%%%%%%%%%%%%%%%%%%%%%%%%%%%%%%%%%%%%%%%
%% Sections
%
\section{Introduction}
    \label{s:Intro}
 
%%%%%%%%%%%%%%%%%%%%%%%%%%%%%%%%%%%%%%%%%%%%%%%%%%%%
% 1. Give a bit more structure (paragraphs etc).
% 2. Link more to work/papers on magnetic flux 
%    distribution
%

% Start with general statements about context and importance of study: dynamo theory (brief), 
%corona and coronal heating, only exact layer with accurate measurements possible.
    
 Improving our understanding of the spatiotemporal behaviour 
 of the distribution of magnetic flux on the solar surface is 
 important for several areas of solar physics. On the one hand, 
 the characteristics of the solar surface magnetic field can provide 
 insights into the dynamo processes taking place in the 
 Sun's interior 
 \citep[e.g.][]{brandenburg2012current,charbonneau2014annual,charboneau2020livrev,brun2017livrev,rempel:etal23}. 
 On the other hand, 
 with accurate measurements of the solar magnetic field currently 
 only possible at photospheric levels
 \citep[e.g.][]{bellotrubio2019livrev,wiegelmann2021livrev}, 
 the structure and dynamics of the photospheric magnetic 
 flux also provides a crucial input into our
 understanding of the structure of the coronal magnetic field 
 and phenomena like coronal heating and magnetic activity processes 
 \citep[e.g][]{mackay2012livrev,petrie2013,pevtsov2021}.
 
 Because it involves the largest scales, the most obvious manifestation of the
 processes generating the solar magnetic field in the solar interior is 
 the 11-year cycle in sunspot numbers, 
 which is a 22-year cycle 
 % TN: revised 15/06/2026
 if 
 the 
 polarity of magnetic fields is taking into account 
 \citep[e.g.][]{solanki2006solar,Hathaway_2015}. While sunspots are
 observed in the active-region bands between $\pm 40^\circ$ latitude, magnetic features of all sizes can
 be observed on the solar surface and smaller magnetic features
 seem to be distributed across the entire solar surface \citep[e.g.][]{solanki2006solar}.
 
  %\begin{itemize}
 
% \item What is the current state-of-the-art?

 Due to their general importance, different aspects of the solar surface magnetic field and
 its characteristics have been investigated in the past. Numerous studies focus on larger magnetic features such as pores, sunspot umbrae, sunspots and sunspot groups. 
 Observations have shown a strong correlation between the magnetic flux and area of sunspots, as well as active regions in general, \citep[e.g.][]{nicholson1933,houtgast1948,pevtsov2014,nagovitsyn2017,pevtsov2021, wang:etal23} and hence the distribution of both of these quantities, as well as active regions in general, has been investigated intensively \citep[e.g.][]{kuklin1980astronomical,bogdan1988distribution,baumann2005size,zharkov2005statistical,schad2010structural,jiang2011solar,nagovitsyn2012possible,cho2015,munoz2015small,kostyuchenko2017,nikbaksh2019,tlatov2019,nagovitsyn2021,sakurai:toriumi23,kumar:etal25}. Other studies have focussed on bipolar magnetic regions \citep[e.g.][]{tang1984statistical,harvey1993properties,zhang2010statistical}, the quiet network magnetic flux \citep[e.g][]{schrijver1997sustaining}, ephemeral regions \citep[e.g.][]{parnell2002nature}, or emerging flux features \citep[e.g.][]{thornton2011small}.
%
% Munoz-Jaramillo separately ??
%

%
% Now discuss magnetic feature studies across all scales:
% Meunier (2003), Parnell et al (2009), Javaherian et al. 
% (2017)
%
Instead of considering either larger scale features or a
specific set of features there have also been a number of studies of the distribution of features across all scales.
For example, \citet{meunier2003statistical} used SOHO/MDI full-disk magnetograms to examine the statistical distribution of the magnetic flux and the variation of this distribution, which they found to be a power law, with the solar cycle. \citet{parnell2009power} combined SOHO/MDI data with observations by Hinode/SOT, which allowed them to cover five orders of magnitude flux. These authors also found that the magnetic flux distribution is represented by a single power law. A more recent study of a similar type,
but concentrating on quiet sun flux elements, has been undertaken by \citet{javaherian2017} on the basis of SDO/HMI magnetograms, with the authors finding a broken (or double) power law distribution function.
%
%
% Discuss different distribution functions used
% and some statistical basis
%
Depending on the data sets and features studied, very different forms
for the distribution functions have been put forward, for example
exponential distributions
\citep[e.g.][]{tang1984statistical,schrijver1997sustaining}, 
log-normal distributions \citep[e.g][]{bogdan1988distribution,baumann2005size,zhang2010statistical,schad2010structural},
double log-normal distributions 
\citep[e.g.]{kuklin1980astronomical,nagovitsyn2012possible,nagovitsyn2021},
or, as already mentioned above, power law distributions 
\citep[e.g.][]{zharkov2005statistical,meunier2003statistical,parnell2009power,thornton2011small,shapoval2018}. 
Other forms of suggested distribution functions are polynomials \citep[e.g][]{harvey1993properties}, the Weibull distribution \citep[e.g.][]{parnell2002nature}, broken (or double) power law distributions \citep[e.g.][]{javaherian2017,song:etal24}, and bi-modal
log-normal functions \citep[e.g.][]{cho2015}, as well as 
combinations of distribution functions such as power law 
and log-normal \citep[e.g.][]{jiang2011solar} and
Weibull and log-normal \citep[e.g.][]{munoz2015small}.
%
% Discuss bi-modality as an indicator for different processes?
%
The distribution functions mentioned above can be split into to 
different classes: unimodal distributions such as, 
for example a power-law distribution, and 
bimodal distributions like the hybrid distributions, 
broken/double power laws or double log-normal distributions. 
Bimodal distributions are often regarded as being indicative
of two different physical processes generating the distinct parts of
the distribution \citep[e.g][]{nagovitsyn2012possible,munoz2015small,song:etal24} or that it could be representing different 
%stage 
%TN revised: 15/06/2026
stages
in the evolution of certain magnetic features \citep[e.g][]{tlatov2019}. On the other hand, 
unimodal distributions \citep[e.g][]{schrijver1997sustaining,parnell2009power} could be 
indicative of a single physical process operating over all scales of the distribution.

 In this paper we carry out a statistical analysis of photospheric magnetic flux features over a full solar cycle 
 using data 
 % TN revised: 15/06/2026
 collected
 by the Helioseismic and Magnetic Imager (HMI) instrument on the Solar Dynamics Observatory (SDO). 
 %
 % New text related to Song et al. (2024)
 %
 The methods and data we use are very similar in general to \citet{song:etal24}. However, a major difference
 is that 
our
    study tests a number of different distributions against our 
%    own 
    data set (more details in Section \ref{s:Data}) to determine the best fitting distribution over a full solar cycle. 
    
    In Section \ref{s:Data} we introduce the data set and our method of analysis before highlighting our results in Section \ref{s:Results}. We discuss the implications of our findings in Section \ref{s:Discussion} and provide some concluding remarks in Section \ref{s:Conclusions}.
    
\section{Data}
    \label{s:Data}

	We use a sequence of 260 full disk line-of-sight magnetograms from the Helioseismic and Magnetic Imager (HMI) on the Solar Dynamics Observatory (SDO) taken at midnight on the 1st and 16th day of each month from 01 May 2010 to 16 March 2021. The dates were chosen arbitrarily but they provide at least half a Carrington rotation (27.27 days) between subsequent observations meaning that for each month we get two completely independent views of the photosphere. Three exceptions to these dates are 16 September 2010, 16 December 2011 and 16 March 2021 where we take the map at 00:12:00 the same day, 23:12:00 the day before and 00:12:00 the same day, respectively. This is due to a lack of good data at midnight of those days. This gives us enough data to span almost a full solar cycle. Magnetograms measure the magnetic flux density (in Mx cm$^{-2}$) on the solar photosphere, however we use the magnetic flux density as a proxy for the average magnetic field strength at each pixel (measured in Gauss, G) Each magnetogram is a 4096$\times$4096 pixel image with a pixel area of 0.5$\times$0.5 arcsec$^2$. The data is corrected from line-of-sight to radial magnetic field strength using a radial cosine correction. Each pixel is multiplied by the area of the Sun which it represents in order to convert the data into a measurement of radial magnetic flux. To avoid inaccuracies in radial magnetic flux near the limb we ignore all pixels further than 60$^\circ$ in latitude or longitude from the center of the Sun.

    \subsection{Feature identification method}
	    \label{s:feat_ID}
	
    	\subsubsection{Comparison of different methods}
	        \label{s:comparison_of_methods}

            In order to identify and isolate individual magnetic features in a magnetogram we use a modified version of the \textit{clumping} algorithm developed by \citet{parnell2002nature}. The \textit{clumping} algorithm defines a magnetic feature as a ``contiguous group of same-signed pixels with absolute values greater than a lower cutoff''. These groups of pixels are known as `flux massifs' as they are analogous to a mountain massif. There are two other common feature detection algorithms \citep{deforest2007solar}, both of which have different definitions of magnetic features. The first of which, known as the \textit{downhill} method \citep{welsch2003magnetic}, finds flux peaks rather than massifs. The algorithm divides massifs into their individual `summits' along saddle lines. The last method, the \textit{curvature} method (\citealp{strous1994dynamics, hagenaar1999dispersal}), finds flux cores by identifying local extrema and grouping together contiguous pixels that form a convex surface around the extrema. Figure \ref{fig:feat_ID} (a) shows a cartoon illustration of the differences in identification of features by the three different methods. We decided to employ the \textit{clumping} algorithm as we feel it is the most robust to changes in the resolution of data being analysed. We decided against the \textit{downhill} and \textit{curvature} methods as they are sensitive to the number of local extrema. Increasing resolution tends to increase the number of individual `peaks' within a flux massif and, hence, the \textit{downhill} and \textit{curvature} methods will identify many smaller features rather than the flux massif itself. We believe the flux massifs are the best representation of individual magnetic features. The \textit{clumping} algorithm holds up well to increases in resolution as it does not split up the features 
%            are 
            present in lower resolution data. Therefore, any newly identified features would be a result of the increased resolution only. In addition, the \textit{curvature} method is disregarded as it also fails to identify any features that have a `flat' top due to their lack of curvature as is shown in Figure \ref{fig:feat_ID}. This may be an uncommon occurrence but is worth taking into account.
        
        \subsubsection{Modification of clumping algorithm}
            \label{s:modification}

\begin{figure}
                \centerline{ \hspace*{-0.03\textwidth}
                 \includegraphics[width=0.5\textwidth,clip=]{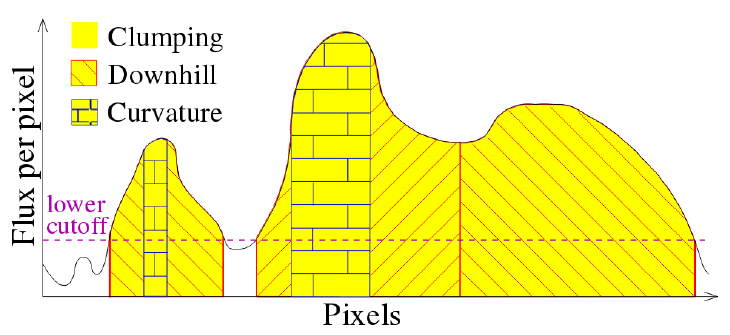} 
                \hspace*{-0.03\textwidth}
                \includegraphics[width=0.5\textwidth,clip=]{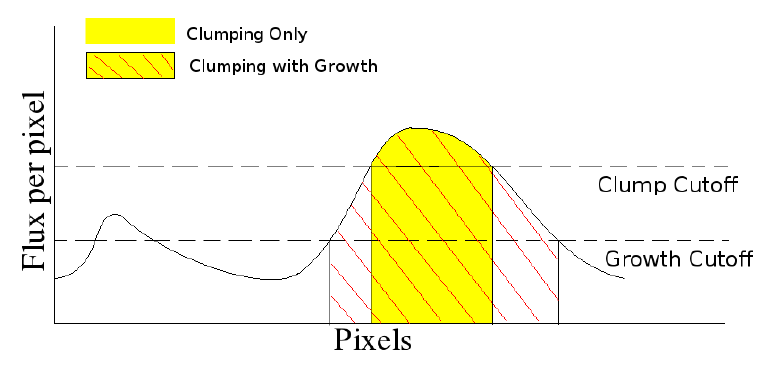}}
                \vspace{-0.25\textwidth}
        \centerline{%\Large %\bf     % Includes the labels (here needs the color 
                                 %   package, see beginning of this file)
           \hspace{0.4 \textwidth}  \color{black}{(a)}
           \hspace{0.45\textwidth}  \color{black}{(b)}
            \hfill}
                \vspace{0.25\textwidth}
                \caption{(a) A cartoon illustration of three common feature identification algorithms. The \textit{clumping} method, 
                % TN revision: 16/06/2026
                as originally devised by \citet{parnell2002nature}, uses a single lower cut-off
                and
                identifies two separate features. The \textit{downhill} method splits the larger feature into two separate features at the local minimum (in 3D this would be a saddle point). The \textit{curvature} method only identifies two small features; the peak at the right hand side is too flat for the method to identify. (b) A cartoon illustration of the filtering process carried out by the separate thresholds. The left hand peak is created by `noisy' pixels and should therefore be neglected. The \textit{clump} cutoff, which is the
                same as the lower cut-off in panel (a), achieves this, however, it removes some pixels which we believe are genuine constituents of the feature at the right hand side. Expanding the feature out to the \textit{growth} cutoff allows us to obtain better estimates of what we believe to be the feature's true size.}
                \label{fig:feat_ID}
            \end{figure}

            We have modified the \textit{clumping} algorithm by allowing for a second threshold (known as the \textit{growth} cutoff) which is lower than the original threshold (known as the \textit{clump} cutoff). The \textit{growth} cutoff allows us to expand the identified features by including any bordering pixels that have an absolute value greater than the \textit{growth} cutoff. We do this so we can identify features more accurately by including all pixels which contribute to the feature and avoid any `false positive' features. We set the \textit{clump} cutoff high enough to avoid identifying `noisy' pixels as features but in doing so we omit any contributing pixels which are below the threshold. The \textit{growth} cutoff allows us to incorporate these pixels into their respective features. The two threshold modification can alternatively be viewed as the \textit{growth} cutoff acting as the original threshold which needs to be satisfied and the \textit{clump} cutoff acting as a filter which tries to remove any `false positives'. Figure \ref{fig:feat_ID} (b) shows a cartoon illustration of this filtering in practise. As a final step we neglect all identified features which do not consist of at least five contiguous pixels.
        
        \subsubsection{Choice of thresholds}
            \label{s:thresholds}
            
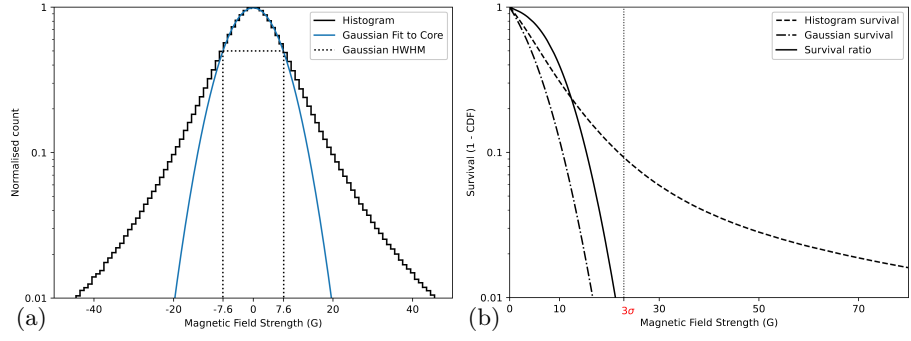
\begin{figure}
        	  \begin{center}
                \includegraphics[width=0.49\textwidth]
                        {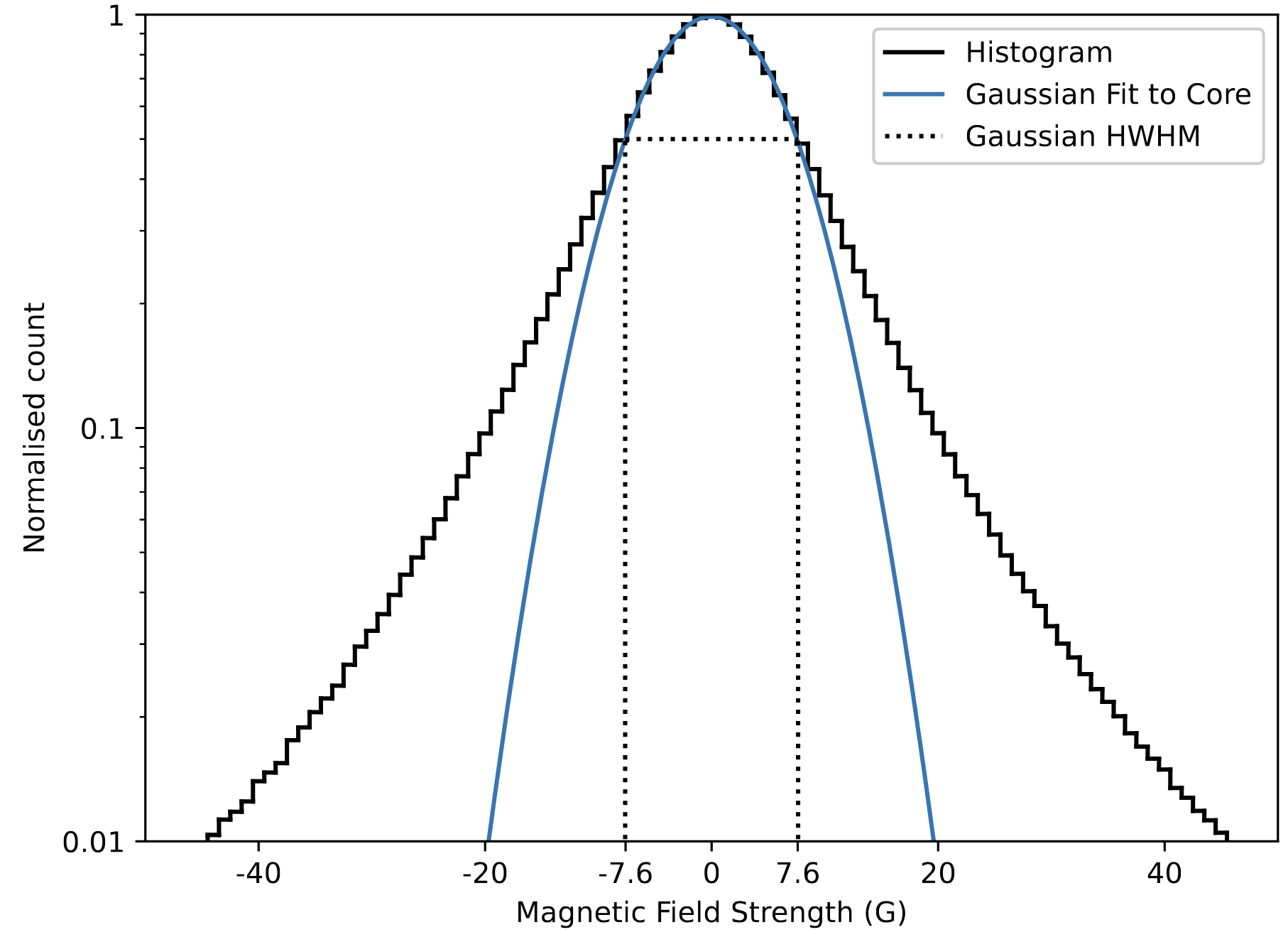}  
                \includegraphics[width=0.485\textwidth]
                        {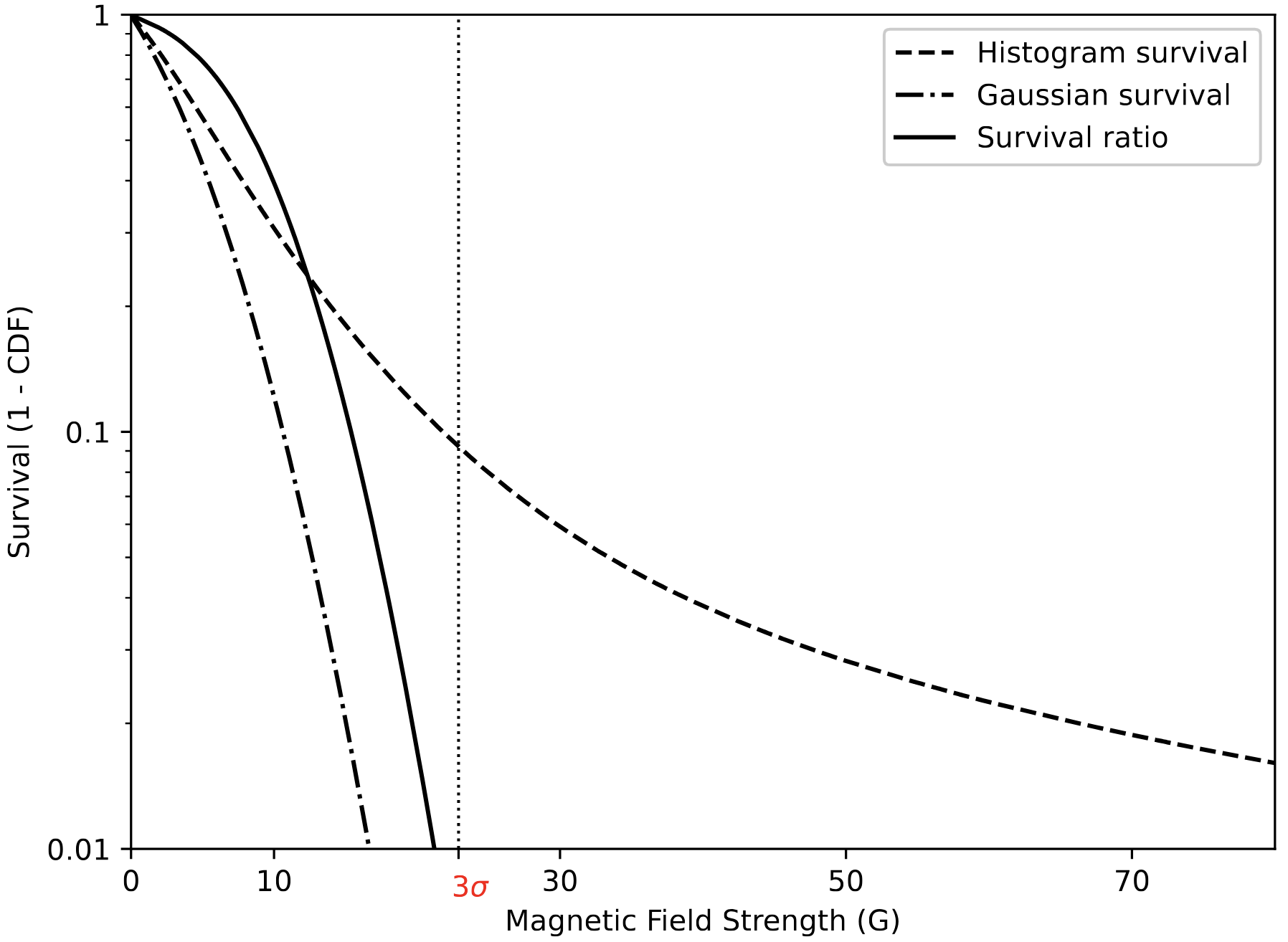}
    \end{center}  
    \setlength{\unitlength}{1cm}
    \begin{picture}(10,0.0)
    %Labels of figure location
    \put(0.25,0.9){\small{(a)}}
    \put(6.23,0.9){\small{(b)}}
    \end{picture}
                \caption{(a) A histogram (black solid line) of the magnetic field 
                strength of the pixels is created. 
                A Gaussian curve (blue solid) is fitted to 
                the core of the histogram (any bins with count greater than or equal to 0.5). The full width at half maximum (FWHM) is also plotted (dotted). The noise of the magnetogram is estimated as the half width at half maximum (HWHM). For this magnetogram the noise level is approximately 
               % 8.131 G. 
               % TN revised: 16/06/2026
               7.6 G.
                (b) The survival function of the histogram (dashed) and the Gaussian curve (dash-dotted) are plotted. The ratio of the two curves (solid) 
                %tells us what 
                indicates the
                fraction of the pixels above a certain magnetic field strength that are associated with noise. From the graph we observe that for a threshold of 3 times the noise level 
                %(24.393 G), 
                %TN revised: 16/06/2026
                (22.8 G),
                less than 1\% of the pixels above the threshold are associated with noise. This allows us to comfortably set a \textit{clump} threshold of 3 times the noise level and a growth threshold of 2 times the noise level to better estimate the true size of the features.}
                \label{fig:noise_analysis}
            \end{figure}

            The two cutoff values are applied as a means to neglect noisy pixels; therefore, we use the noise levels of the magnetograms themselves to assign the cutoff values. We estimate the noise level of the magnetograms following the technique of \citet{parnell2002nature}. We fit a Gaussian curve to the core of a histogram of the pixel values assuming that the core of the histogram is associated with the noisy pixels. The range of the noisy pixels is then given as the 
            full-width at half-maximum (FWHM) of the Gaussian fit and the noise level 
            assumed to be the half-width at half-maximum (this accounts for the sign of the individual pixels). 
            To
            give an illustration of the method, Figure \ref{fig:noise_analysis} 
            (a) shows the histogram of the magnetogram taken on 01 May 2014 with 
            the Gaussian fit superimposed. Also shown is the FWHM (dotted). 
            In this case the noise level is approximately 8.131 G. 
            To determine the thresholds we perform a survival analysis; 
            a statistical technique which can be used to determine the 
            proportion of a population which will survive past a certain time. 
            In our case we can use survival analysis to determine the 
            fraction of pixels which are associated with noise. 
            Figure \ref{fig:noise_analysis} (b) shows the survival function for 
            both the histogram and the fitted Gaussian (dashed and dash-dotted, respectively). 
            The survival function is defined as $1 - F(x)$, where $F(x)$ is the 
            cumulative distribution function for the observed distribution. 
            The ratio of the survival functions of both curves (solid) tells us what 
            fraction of pixels with field strength above a certain 
            value are associated with noise. From the figure we see that 
            for pixels with magnetic field strength greater than 3 times the noise 
            of the magnetogram (24.393 G) less than 1\% of these are associated with noise. 
            We therefore choose the \textit{clump} threshold to be 3 times the noise level 
            and allow for growth of features back to a \textit{growth} threshold set at 2 
            times the noise level. However, the noise level is different for every magnetogram 
            and we want to ensure uniformity of feature identification across the full solar cycle. 
        
\begin{figure}
                 \centerline{ 
                \includegraphics[width=\textwidth,clip=]{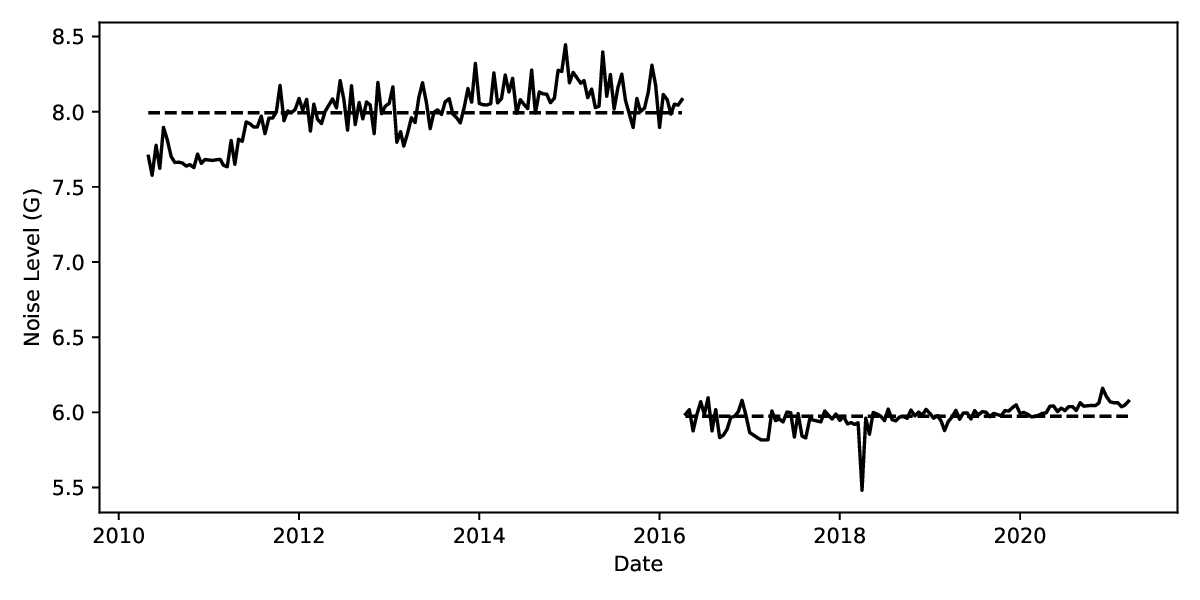}
                }
                \caption{The noise level across the full data set. The drop in noise level is caused by a change in the way HMI takes measurements \citep{liu2016hmi} which reduced the noise associated with magnetograms. The dashed lines show the average value of the noise level before and after the drop; approximately 7.993 G and 5.974 G, respectively. There is no obvious correlation between noise level and solar cycle so we use the average noise values as the thresholds for all maps to ensure uniformity in the feature detection.}
                \label{fig:noise_cycle}
            \end{figure}

            Figure \ref{fig:noise_cycle} shows the value of noise for each magnetogram in our data set. The first thing to note is the significant drop in noise level during April 2016. This is caused by a change in HMI's observing scheme, known as ``Mod-L'', to now combine measurements from two separate cameras. The result of this change is that the noise of magnetograms from HMI is significantly reduced \citep[see][for more details]{liu2016hmi}. The average values of the noise before and after the switch are approximately 7.993 G and 5.974 G, respectively. There is no significant change in the noise levels over time, therefore we can assume that using the average noise value before and after the drop as a proxy for the true noise is acceptable and provides uniformity across all maps. We then calculate the two thresholds, \textit{clump} and \textit{growth}, by multiplying the average noise level by 3 and 2, respectively. We apply these thresholds to the modified clumping method outlined in Section \ref{s:modification} and determine all magnetic features present on each magnetogram.
        
\section{Results}
    \label{s:Results}

\begin{figure}
	\centerline{
    \includegraphics[width=0.5\textwidth,clip=]{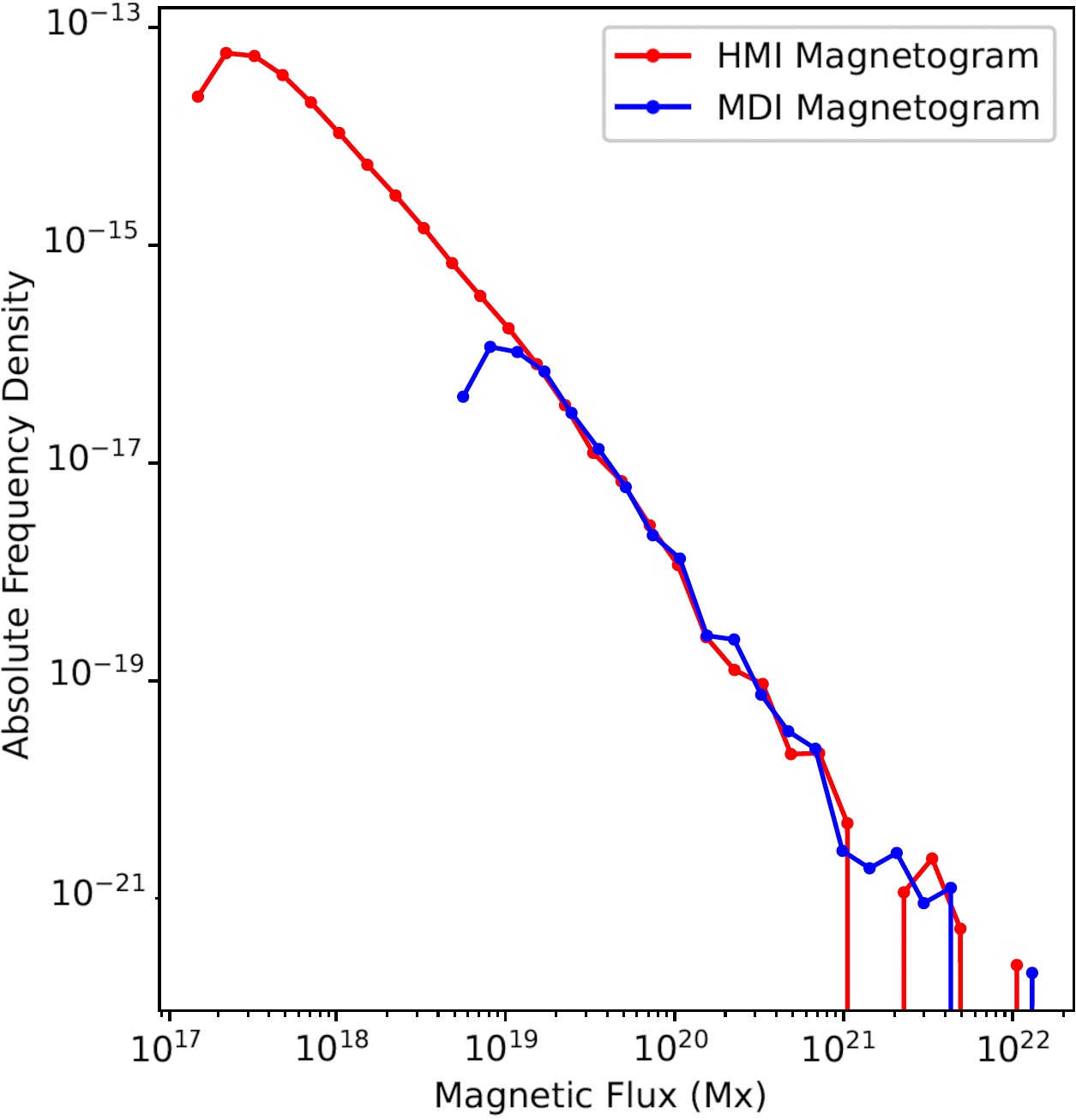}}
    \vspace{-0.115\textwidth}
    %\centerline{ %\large %\bf 
    %            \hspace{0.085 \textwidth} \color{black}{(a)}
    %            \hspace{0.41\textwidth}  (b)
    %            \hfill }
    \vspace{0.13\textwidth}           
	\caption{%(a) Histogram of 
 %absolute 
 %frequency density of magnetic flux features detected in SDO HMI magnetogram taken at midnight on 16 March 2011. (b) comparison of the SDO HMI histogram with a SOHO MDI histogram taken at approximately the same time. 
 %(c) HMI histogram after truncating data to retain only features which have magnetic flux greater than $10^{18}$ Mx.
 Histogram of frequency density of magnetic flux features detected in SDO HMI magnetogram taken at midnight on 16 March 2011 (red) and detected in a SOHO MDI magnetogram taken at approximately the same time (blue). 
 }
	\label{fig:hist_mdi_comp}
\end{figure}

%
% TN 21/11/2022: Revision plan
% 1. Discuss pre-selection of PDFs, reducing to 4
% 2. Rewrite text more according to CNN thesis
%
Following the example of previous work \citep[e.g.][]{parnell2009power}, we use histograms to
analyse the observed distribution of magnetic flux features. Figure \ref{fig:hist_mdi_comp} (a)
is an example for such a histogram. It shows the absolute frequency density of 
magnetic flux for magnetogram data taken on 16 March 2011, plotted on double-logarithmic axes. This example histogram highlights some general properties of magnetic flux histograms.

% The first step towards understanding the observed distribution of fluxes is to look at a histogram. Figure \ref{fig:hist_mdi_comp} (a) shows the histogram of absolute frequency density of magnetic flux features for the 16 March 2011 magnetogram plotted on doubly-logarithmic axes. We notice that there is a drop-off in frequency at the lower flux tail of the distribution. This drop-off is caused by an inability to resolve the smallest features correctly and, therefore, a lot of smaller features are neglected by the feature identification algorithm \citep{parnell2009power}. 

Firstly, over a considerable range of magnetic flux ($10^{18} - 10^{20}$ Mx) the distribution seems to 
be well approximated by a straight line. In a double-logarithmic plot this is interpreted as an
indicator of a power law distribution \citep[e.g.][see the Appendix A for mathematical details]{newman2005power,parnell2009power}. 

Secondly, however, one notices that above $10^{20}$ Mx the data becomes sparser and the straight-line behaviour is less evident. 
It is one of the aims of this paper to investigate the nature of 
the magnetic flux distribution over the full range of flux values; 
another aim is to find out whether and how the distribution varies over a solar cycle.

Thirdly, we notice a turnover in the lower flux tail ($< 10^{18}$ Mx). For context we show in panel (b) of
Figure \ref{fig:hist_mdi_comp}, a comparison between the SDO HMI data and a SOHO MDI observation taken 
at the same time 
\citep[for details how to account for the difference in 
measurement between MDI and HMI, see][]{liu2016hmi}.
As one can see in Figure \ref{fig:hist_mdi_comp}(b) the MDI based histogram has a turnover
at a higher magnetic flux value due to the lower resolution of MDI compared to HMI, while otherwise
the histograms seem to match closely. Although this is not a proof, we take this as an indication that the turnover in the HMI histograms is also caused by the limitation in the 
observational resolution of the instrument.

% Panel (b) helps to emphasise the artificial nature of the drop-off. The histogram of features identified in an HMI magnetogram are compared to those detected in a magnetogram taken by the Michelson Doppler Imager (MDI) at approximately the same time. It is noted that both histograms exhibit a drop-off albeit at different values of magnetic flux. The fact that the drop-offs occur at different values is an indication that they must be artificially caused by the separate instruments' observational limits. MDI and HMI have different spatial resolutions; HMI having a higher resolution means features of lower flux can be detected more accurately, thus leading to a drop-off at a lower value. 

As the drop-off 
%is 
seems to be
%artificial 
caused by instrumental effects,
we have to 
%neglect 
exclude
it if we are to obtain the true underlying distribution of magnetic flux features. 
To account for this we truncate the dataset to only include features 
with a magnetic flux greater than $10^{18}$ Mx and re-bin the histogram.
%, as is shown in panel (c). 
This limit is chosen somewhat arbitrarily but setting a hard limit is a more sensible approach than trying to determine separate limits on a case-by-case basis.

% Rewrite next paragraph slightly

% Post-truncation, the histogram data somewhat resembles a straight line on doubly-logarithmic axes. Such behaviour  is often taken to be indicative of a power law probability 
%%distribution 
% density
%function \citep[e.g.][]{newman2005power}, but in this paper we will
% conduct a more careful investigation by testing
% a number of different PDF models 

% We can use this to guide our approach to formulating a model which may represent the observed data, as straight line behaviour on doubly logarithmic axes is indicative of a power law model \citep{newman2005power}.

% 1. Take the following single power law discussion out
% 2. Connect directly to selection of PDFs
% 3. After PDF introduction, make pre-selection using P-P plots?

% New text added by TN 30 /07/2024

\begin{figure}
	\begin{center}
   \includegraphics[width=0.465\textwidth,clip=]
    {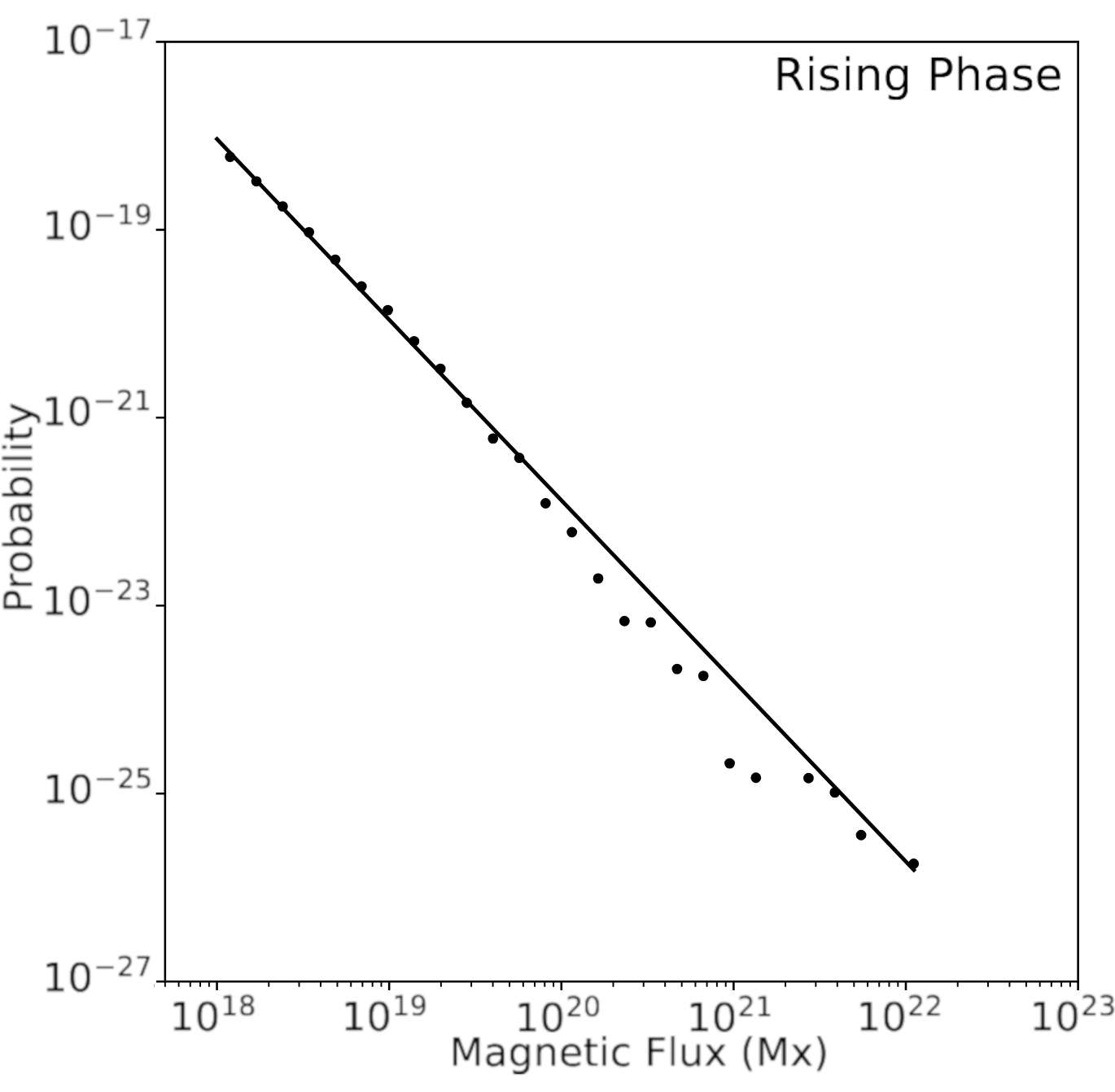}  
    \includegraphics[width=0.465\textwidth,clip=]
    {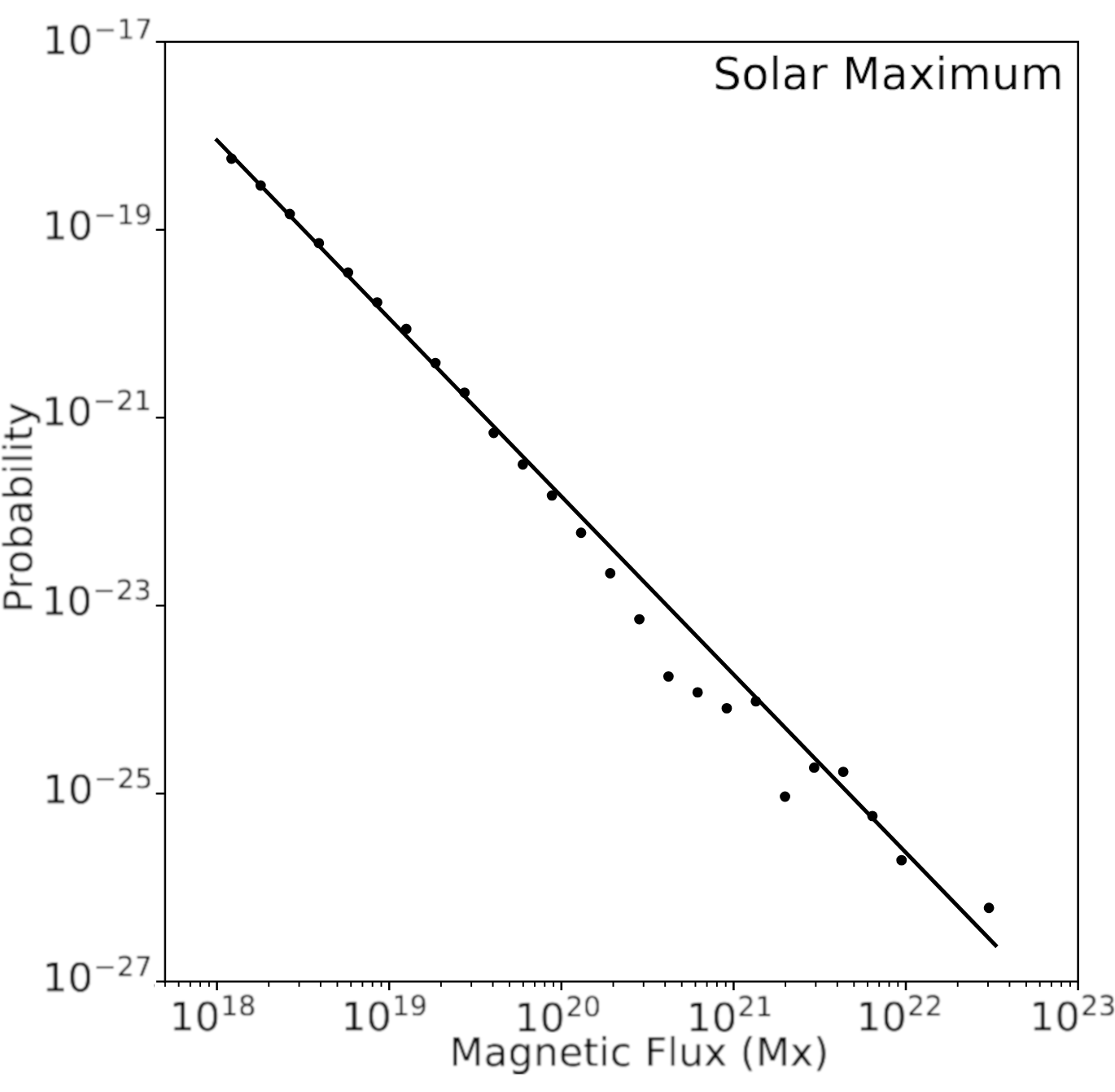}
    \includegraphics[width=0.465\textwidth,clip=]
    {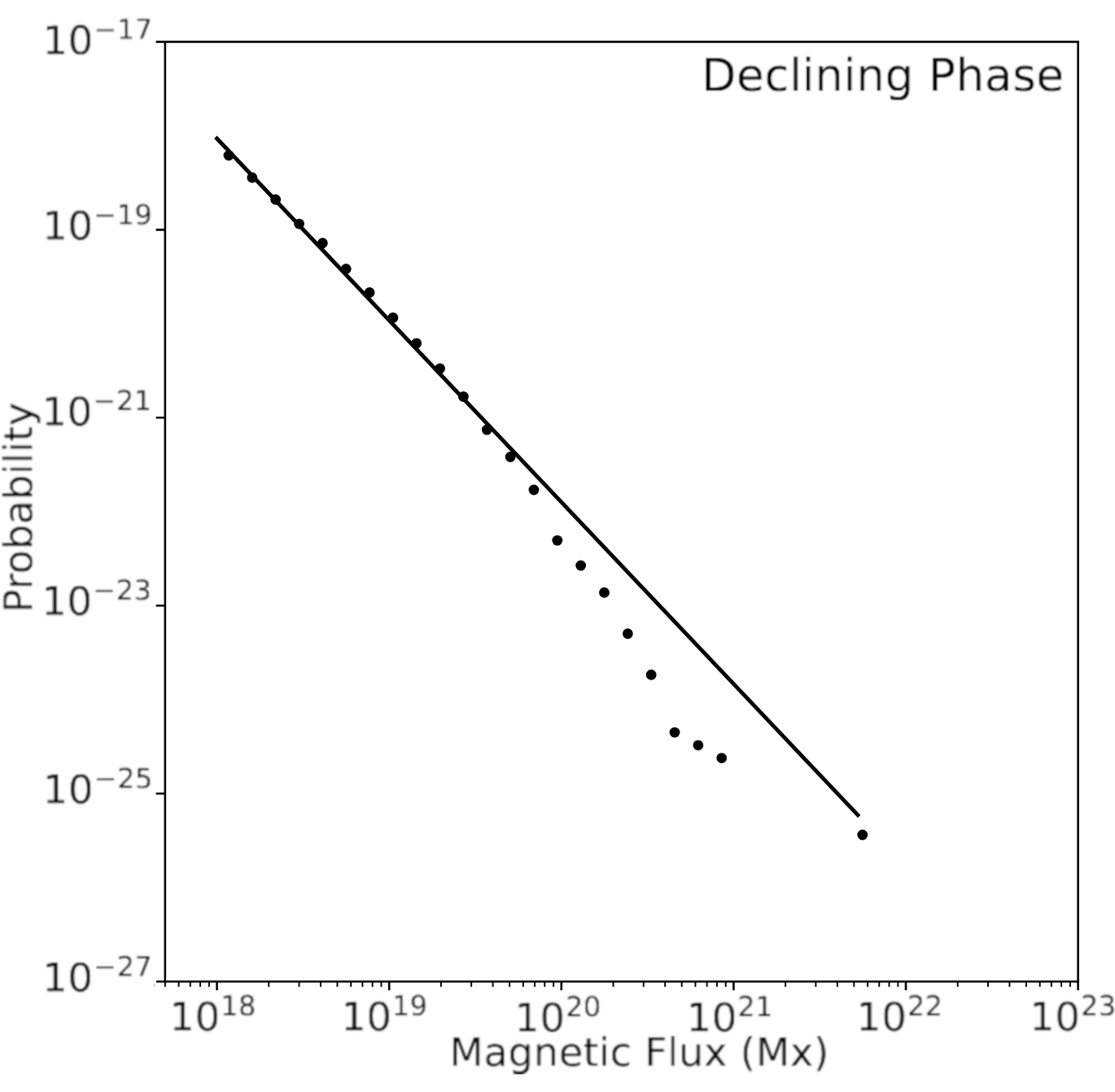}
    \includegraphics[width=0.465\textwidth,clip=]
    {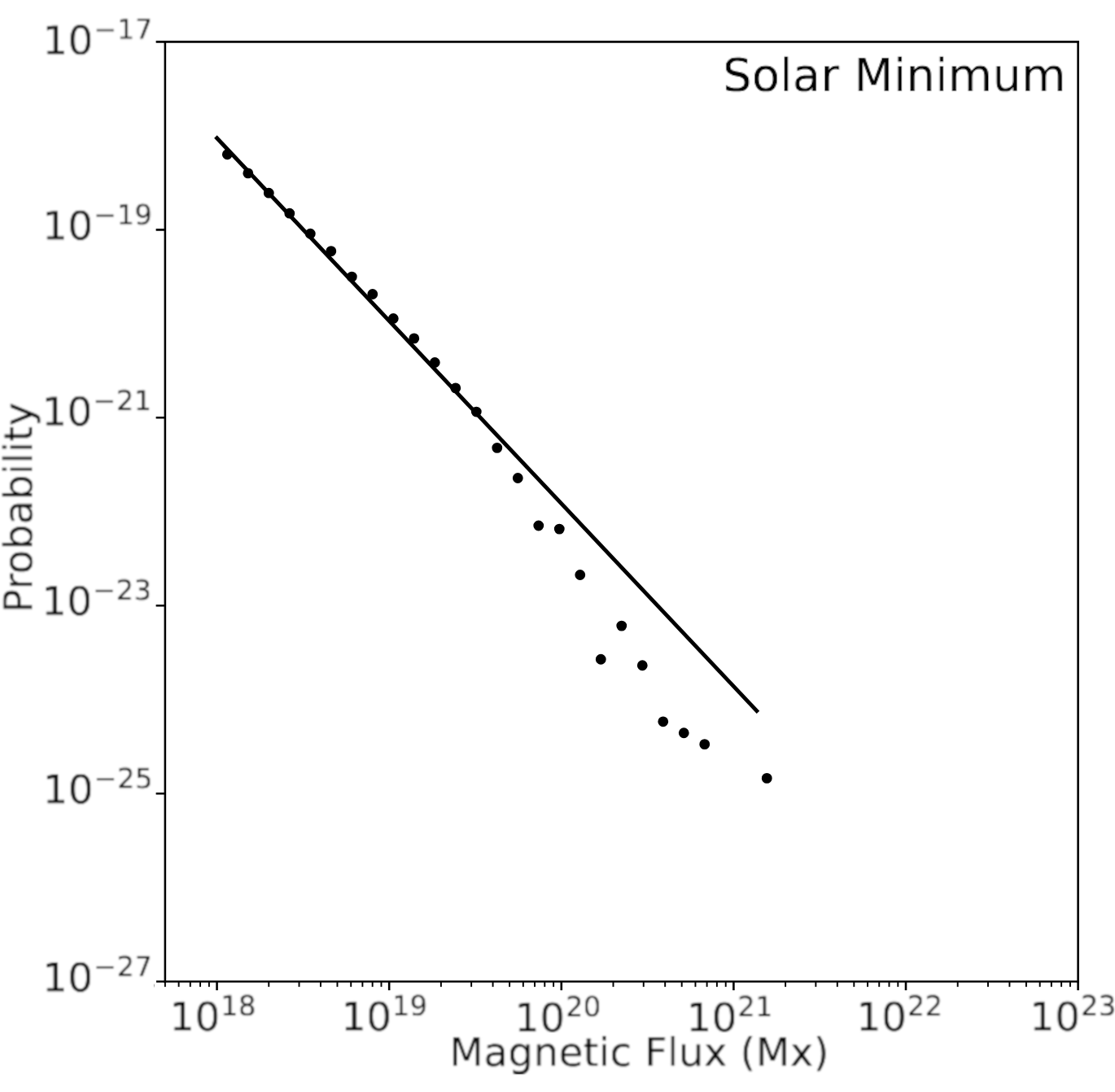}
    \end{center}  
    \setlength{\unitlength}{1cm}
    \begin{picture}(10,0.0)
    %Labels of figure location
    \put(0.65,6.53){{(a)}}
    \put(6.35,6.53){{(b)}}
    \put(0.65,0.9){{(c)}}
    \put(6.35,0.9){{(d)}}
    \end{picture}
	\caption{Histogram of magnetic flux features and a fitted power law model at different times in the solar cycle. (a) 16 March 2011 representing rising activity. (b) 01 July 2014 representing solar maximum. (c) 16 August 2017 representing declining activity. (d) 16 August 2020 representing solar minimum.}
	\label{fig:single_power_hists}
%\vspace{-0.46\textwidth}   % Shift close to the panel top 
%\centerline{%\Large \bf     % Includes the labels (here needs the color 
                          %   package, see beginning of this file)
%\hspace{0.43 \textwidth}  \color{black}{(a)}
%\hspace{0.43\textwidth}  \color{black}{(b)}
%\hfill}
%\hspace{0.43 \textwidth}  \color{black}{(c)}
%\hspace{0.43\textwidth}  \color{black}{(d)}
%   \hfill}
\end{figure}
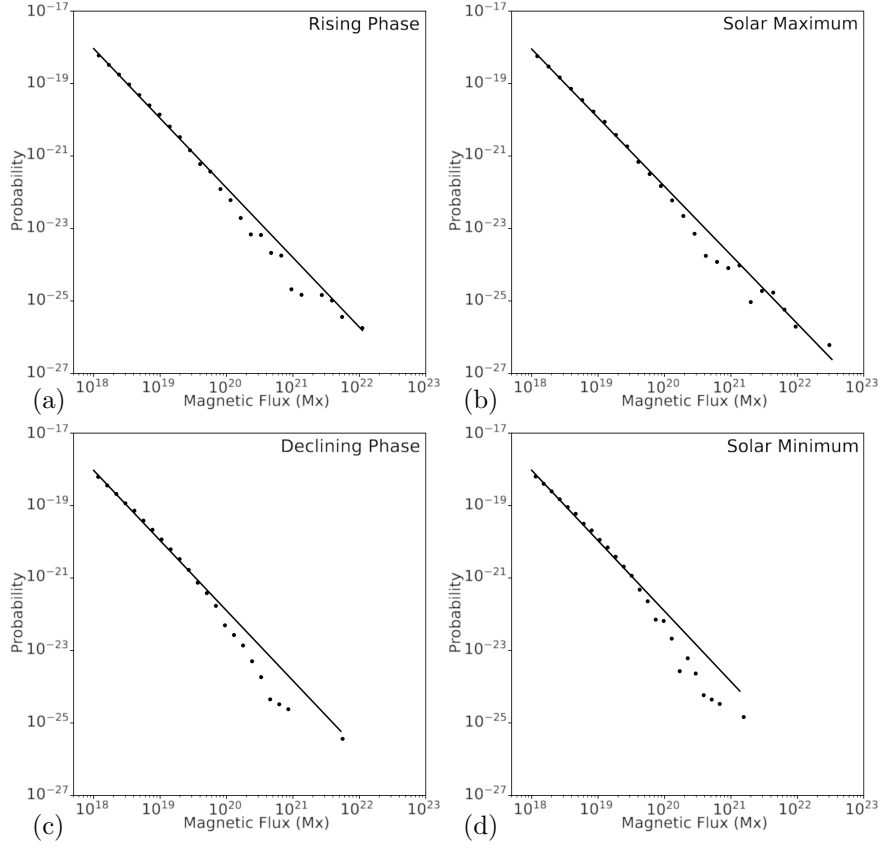

As already mentioned before, one notices that the truncated histogram data resemble a straight 
line in a double-logarithmic plot. This is usually interpreted as being indicative of an underlying 
power distribution function \citep[e.g.][]{newman2005power,parnell2009power}. 
Figure \ref{fig:single_power_hists} shows 
%the histogram 
histograms of the observed distribution of magnetic flux at different phases of the solar cycle, together with a 
power law model that has been fitted to the 
%TN revised: 15/06/2026
%date 
data
using the maximum likelihood method 
\citep[e.g.][]{scholz2014maximum}.

One can see that the 
%single 
power law 
%performs 
fits the data reasonably well in each case. However, it is also clear that in the 
higher flux tail the power law model starts to deviate from the observed distribution. 
The histogram values tend to be lower than the power law model; this 
is especially true during the declining phase and solar minimum. 
Therefore the question arises whether a power law is really the 
optimal mathematical model to represent the data or whether other distribution 
functions might represent the data better.

%Histograms can give a reasonable idea of what sort of model the observed data follow but they only provide a 
We have picked a (single) power law based on a purely subjective interpretation of the data.
Clearly,
%More 
more sophisticated means are needed to make an objective decision 
on whether or not a mathematical model is an accurate 
representation of the true distribution that the data come from. 
In this paper we use a number of different methods to achieve this.

We also we want to investigate whether there might be 
alternative models which perform better than the single 
power law over the full solar cycle. Therefore, in addition to 
the single power law, we will test
eight 
%six 
different 
%models 
probability density functions (PDFs) against the data:
\begin{itemize}
%    \item Single Power Law PDF
    \item Exponential PDF
    \item Lognormal PDF
    \item Weibull PDF
    \item Truncated Weibull PDF (abbreviated as {\em Tr.~Weibull} from now on)
    \item Sharp Double Power Law PDF (abbreviated as {\em Sharp-DPL} from now on)
    \item Smooth Double Power Law PDF (abbreviated as {\em Smooth-DPL} from now on)
    \item Weibull-Lognormal PDF
    \item Truncated Weibull-Lognormal PDF (abbreviated as {\em Tr.~Weibull-Lognormal} from now on)
\end{itemize}
The mathematical details for each of these distributions can be found in Appendix A. We remark that apart from the single power law 
%some 
a few more
of these  
distribution 
functions (exponential, lognormal, and Weibull) have been discussed in the context of magnetic flux distributions before \citep[see e.g.][]{munoz2015small}.

% Need to put p-p plots in
% Link to p- values and exclude Exponential, Lognormal and Weibull
% Show only one of the full cycle p-values as example, the other 
% two are very similar and always below 0.1 ==> exclude

\begin{figure}
\centerline{
\includegraphics[width=1.0\textwidth,clip=]{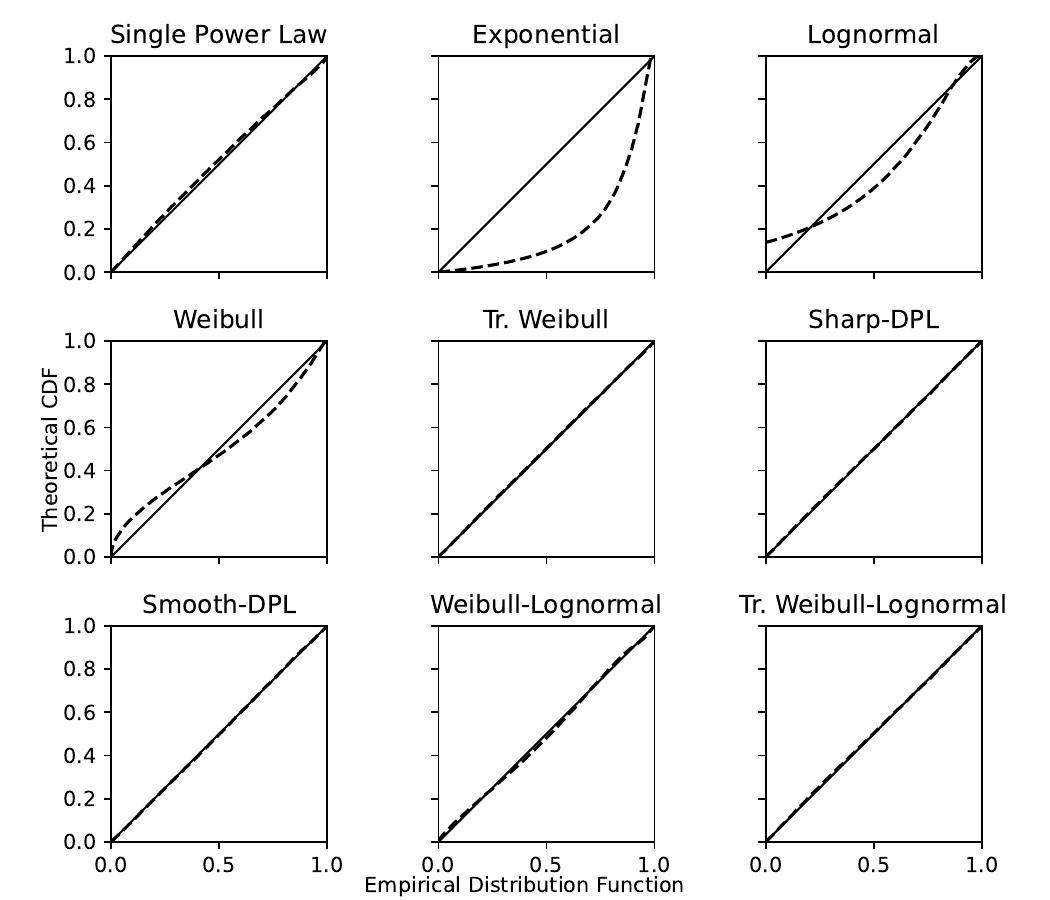}}
\caption{Example P-P plots for the nine model PDFs considered in the paper. 
These plots use data from 1 July 2014, which is around solar maximum.
The solid line in each plot is the main diagonal and the dashed line is the curve
generated by plotting the theoretical CDF against the empiral distribution function.
See main text for more details.}
\label{fig:p-p-plots}
\end{figure}
We start our investigation of these model PDFs by first using a better graphical representation to assess how well a model distribution function 
fits the data called probability-probability plots (P-P plot). A P-P plot 
displays a model's cumulative distribution function (CDF) against the empirical 
cumulative distribution function \citep[e.g.][]{gibbons2020nonparametric} on
the unit square. A good fit between the model CDF and the empirical CDF would 
lead to a curve very close to the main diagonal. A typical example of P-P plots 
for all nine model distribution functions is shown in Fig. \ref{fig:p-p-plots} 
for a date around solar maximum (1 July 2014). From Fig. \ref{fig:p-p-plots} it 
is clear that the exponential, the lognormal, and the Weibull distribution do
not seem to represent the data as well as the other six distribution functions.
However, although P-P plots give a better impression of the quality of 
fit for a model probability 
%distribution 
density
function, the results are still subjective. Furthermore, applying this method to all datasets in our sample would not be practical.

\cite{clauset2009power} suggest that one way to check if a model is suitable or not is to determine the model's `plausibility' by generating a $p$-value based on some goodness-of-fit statistic. We follow their approach by generating the so-called $p$-value based on the Kolmogorov-Smirnov 
statistic (KS Stat). 
The KS Stat is a measure of the maximum difference between a model's cumulative distribution function and the observed data's empirical distribution function \citep[e.g.][]{chakravarti:etal1967}. 

The $p$-value is determined by fitting a model to the data and calculating the KS Stat for the model. Synthetic datasets are then generated based upon the fitted model's parameters, before refitting the model to the synthetic dataset and calculating a new KS Stat. The $p$-value is the fraction of data sets that have a larger KS Stat than the original data. Again, following the example of \cite{clauset2009power}, a model is rejected as implausible if the $p$-value is below $0.1$. 

\begin{table}
	\caption{$p$-values for the plausibility of all nine models for 1 July 2014. Models which surpass the 0.1 significance level are determined to be plausible fits to the data and are highlighted in bold.}
	\begin{tabular}{|c|c|} 
		\hline
		Distribution & $p$-value \\ 
		\hline
		\textbf{Single Power Law} & $\mathbf{0.4}$ \\ 
		Exponential & 0 \\
		Lognormal & 0 \\
		Weibull & 0 \\
		\textbf{Tr. Weibull} & \textbf{1} \\
		\textbf{Sharp-DPL} & \textbf{1} \\
		\textbf{Smooth-DPL} & \textbf{1} \\
		\textbf{Weibull-Lognormal} & $\mathbf{0.2}$ \\
		\textbf{Tr. Weibull-Lognormal} & \textbf{1} \\  
		\hline
	\end{tabular}
	\label{tab:maximum_p_val_table}
\end{table}

In Table \ref{tab:maximum_p_val_table} we provide the $p$-values for all nine 
model probability 
%distribution 
density
functions for the same data used in the P-P 
plots in Fig. \ref{fig:p-p-plots}. One can see that the $p$-values for the Exponential, Lognormal, and Weibull distributions are zero and hence below the 
threshold values of $0.1$, which corroborates the conclusion that these three PDFs do not represent the data well and should be rejected. 

Obviously, one should not base the rejection of a probability 
%distribution 
density
function on a comparison with a single data set. Therefore we have calculated the $p$-values for all models over the complete data set covering the full solar cycle.

It turns out that the $p$-values of the Exponential, 
Lognormal and Weibull PDFs are very close to zero 
throughout the complete data set. 
Plots of the $p$-values for these PDF models 
against time do not convey any additional 
information and are hence omitted. 
On the basis of this result we will 
discard these three PDF models from the analysis.

\begin{figure}
	\centering
    \includegraphics[width=0.8\textwidth]{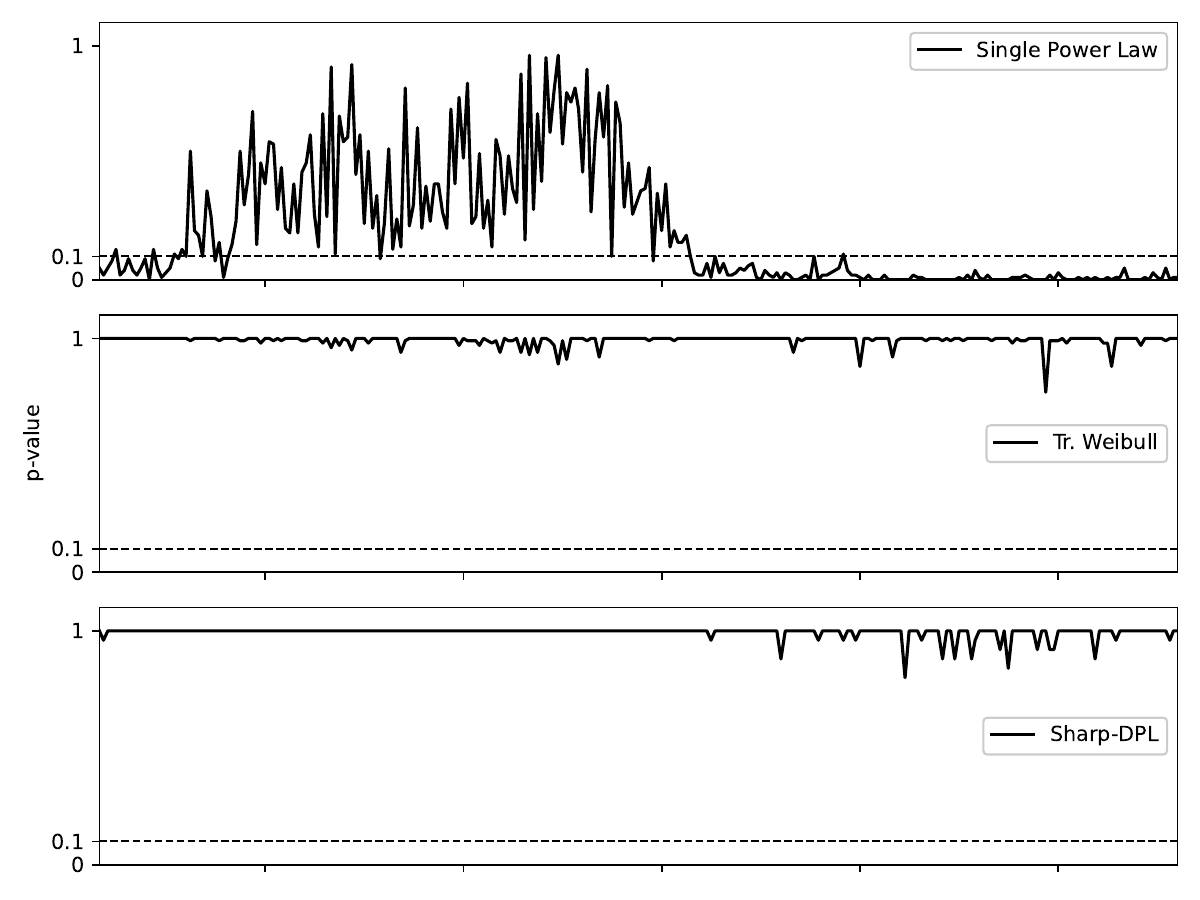}
	\includegraphics[width=0.8\textwidth]{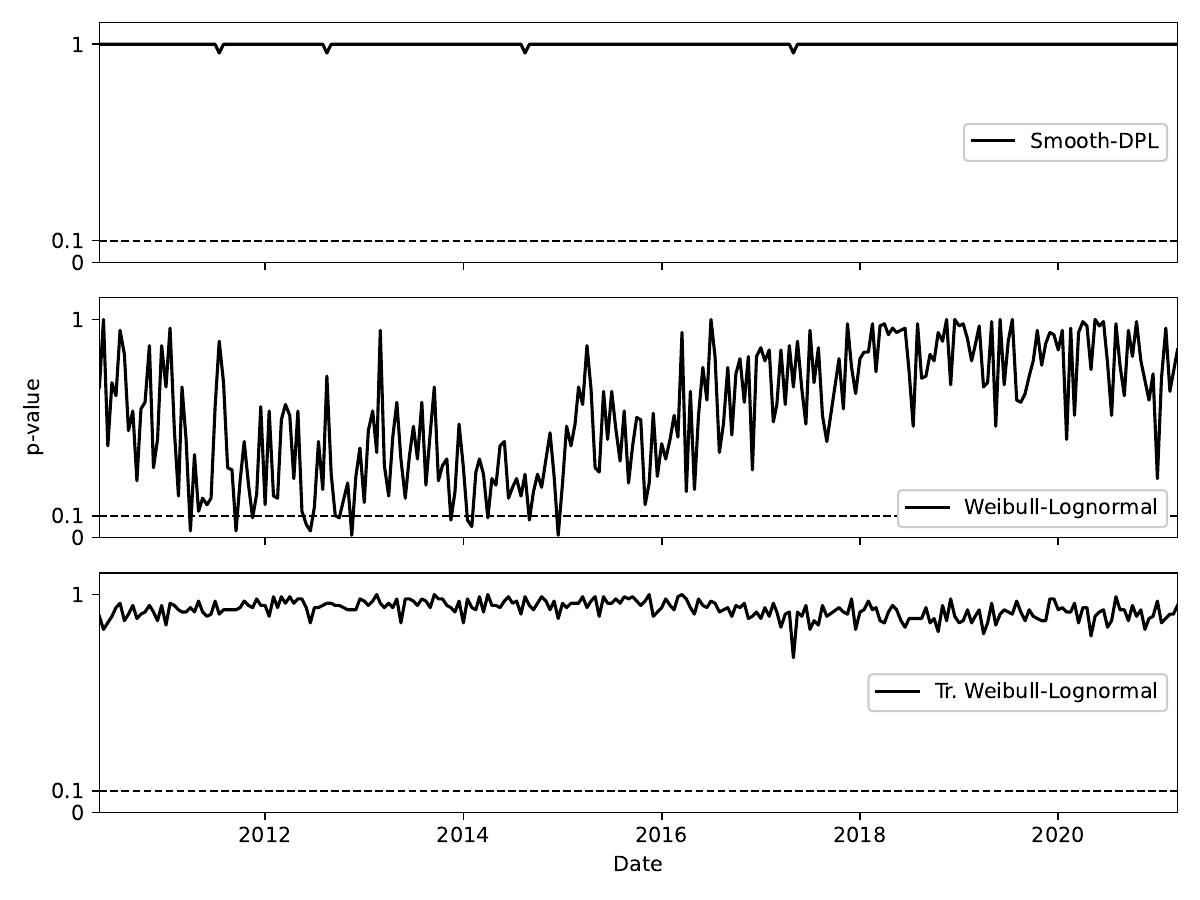}
	\caption{Variation of $p$-values for the remaining
 six PDF models over the full solar cycle. The dashed line 
% represents 
%TN revised: 15/06/2026
shows
 the $0.1$ significance level which plausible models must exceed.
 % Plausibility of PDF models over the full solar cycle based on $p$-value.
 }
	\label{fig:cycle_p_val}
\end{figure}

Figure \ref{fig:cycle_p_val} shows the temporal 
variation over the solar cycle of the $p$-value 
for each of the remaining
six models. In each plot, the 0.1 significance level 
is shown as a dashed line. For plausibility, 
a model's $p$-value must exceed 0.1. 

For the Truncated Weibull, Sharp Double Power Law, 
Smooth Double Power Law and 
Truncated Weibull-Lognormal model PDFs, the $p$-value
is safely above the 0.1 threshold for all magnetograms 
in the sequence. We conclude that each of these models
are plausible fits to the magnetic flux distribution
at all times in the cycle. 

However, both the Single Power Law and 
Weibull-Lognormal models drop below 0.1 on occasion. 
Interestingly, the Single Power Law's $p$-value shows
correlation with the solar cycle, increasing as 
magnetic activity increases, and the Weibull-Lognormal 
model appears to be anti-correlated, decreasing as 
magnetic activity increases. This suggests that the 
Single Power Law may be a good fit at solar maximum 
but fails at solar minimum and likewise, the 
Weibull-Lognormal may be a good fit at solar maximum 
but may fail at solar minimum.

We therefore omit both the single power law and the 
Weibull-Lognormal PDFs from the further analysis and 
accept the other four PDFs as plausible fits to the data.

To be able to make a choice which of the remaining four PDF models is the most suitable, we compare a number of different goodness-of-fit tests 
across the full solar cycle. We have based our choice of goodness-of-fit test statistics on those employed by \cite{munoz2015small}; namely the KS Stat, the relative Akaike Information Criterion (Rel. AIC), and the Akaike weights ($A_w$). However, we have added a fourth test which measures the relative log-likelihood of each model (Rel. Log-Lik). 

The goodness-of-fit statistics are summarised 
in Tables \ref{tab:gof_avg_value_table} and \ref{tab:gof_win_percentage_table} which 
show the average value, over the full solar cycle, of each test statistic for all four models, 
and the percentage of time that each model wins a particular 
goodness-of-fit comparison, respectively. It is clear from these tables 
that the smooth double power law is the best performing model for most of the solar cycle. 
It is difficult to 
decide
whether 
the sharp double power law PDF or the truncated Weibull-lognormal PDF follow in second place based on the win percentages alone, but the truncated Weibull-lognormal PDF comes out ahead on average values. Finally, it is clear that the truncated Weibull PDF is the worst performer over the full cycle. 

\begin{table}
	\caption{The average value, over the full solar cycle, of the four goodness-of-fit statistics for each model.}\label{tab:gof_avg_value_table}
	\begin{tabular}{ccccc}
		\hline
		Model & \multicolumn{4}{c}{Average Value} \\
		& KS Stat & Rel. AIC & $A_w$ & Rel. Log-Lik. \\ 
		\hline
		Tr. Weibull & 0.0073 & 19.532 & 0.036 & $-11.804$ \\
		Tr. Weibull-Lognormal & 0.0062 & 7.469 & 0.176 & $-2.772$ \\
		Sharp-DPL & 0.0082 & 10.497 & 0.211 & $-6.286$ \\
		Smooth-DPL & 0.0055 & 1.966 & 0.577 & $-1.020$ \\		
		\hline
	\end{tabular}
\end{table}

\begin{table}
	\caption{The amount of time each model wins the comparison of each goodness-of-fit statistic, given as a percentage of all magnetograms.}\label{tab:gof_win_percentage_table}
	\begin{tabular}{ccccc}
		\hline
		Model & \multicolumn{4}{c}{Win Percentage (\%)} \\
		& KS Stat & Rel. AIC & $A_w$ & Rel. Log-Lik. \\ 
		\hline
		Tr. Weibull & 20.00 & 2.69 & 2.69 & 0.00\\
		Tr. Weibull-Lognormal & 25.77 & 13.07 & 13.07 & 26.54 \\
		Sharp-DPL & 6.54 & 21.92 & 21.92 & 6.92 \\
		Smooth-DPL & 47.69 & 62.31 & 62.31 & 66.54 \\		
		\hline
	\end{tabular}
\end{table}

Although the smooth double power law does not win 
all goodness-of-fit comparisons 100\% of the time, it 
seems
sensible 
to 
select the
model 
which
performs
best on average. All four models are plausible fits to the data, 
but we wish to make a definitive choice in order to 
perform statistical inference. Inferences about the data may 
come in a number of different forms, for example, using the 
model to make future predictions of the long-term trends in the data; 
to provide a physical constraint in numerical models of the 
sub-surface processes; or to achieve a better theoretical 
understanding of the underlying solar dynamo processes.

\begin{figure}
	\begin{center}
   \includegraphics[width=0.465\textwidth,clip=]
    {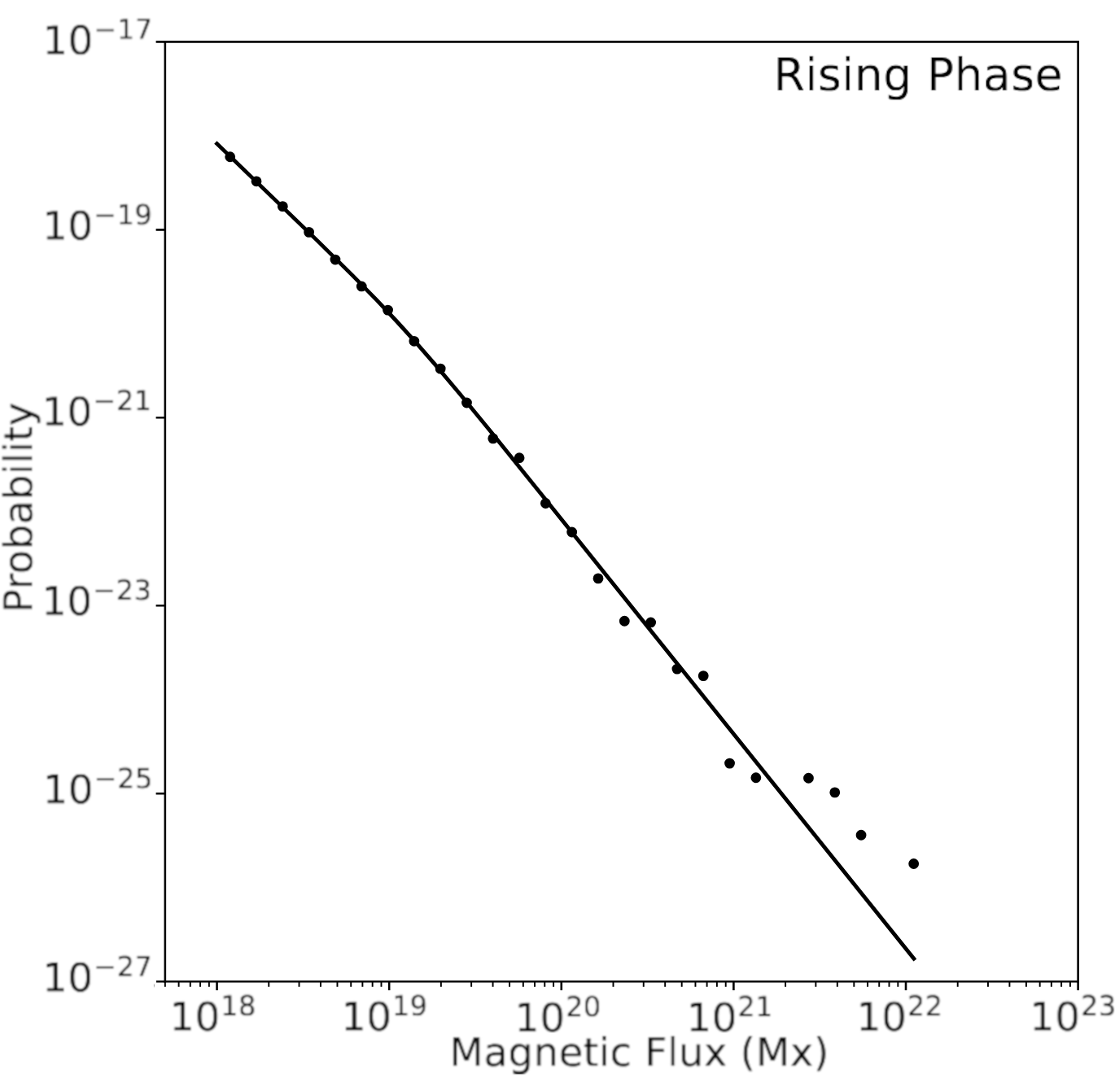}  
    \includegraphics[width=0.465\textwidth,clip=]
    {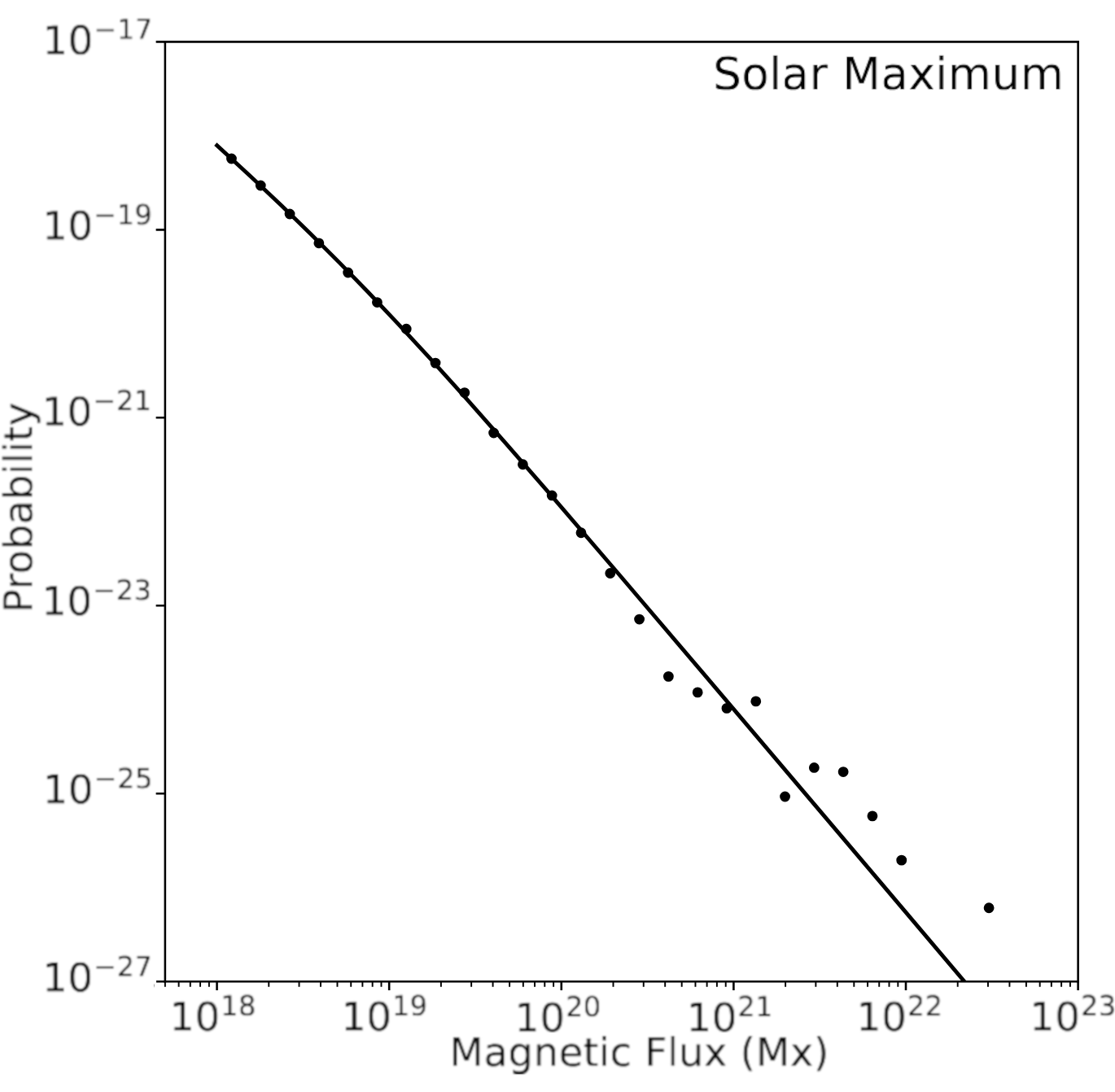}
    \includegraphics[width=0.465\textwidth,clip=]
    {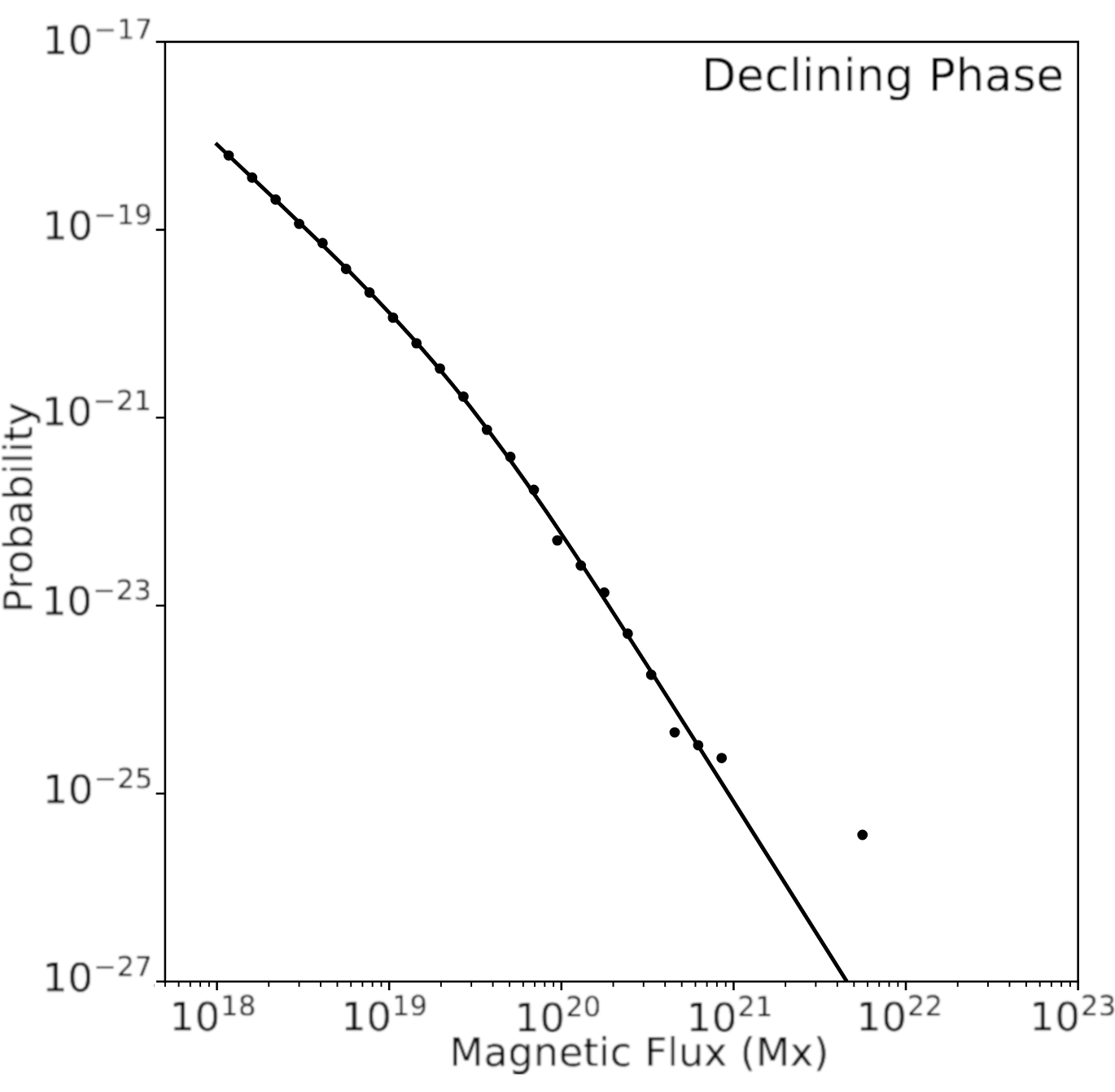}
    \includegraphics[width=0.465\textwidth,clip=]
    {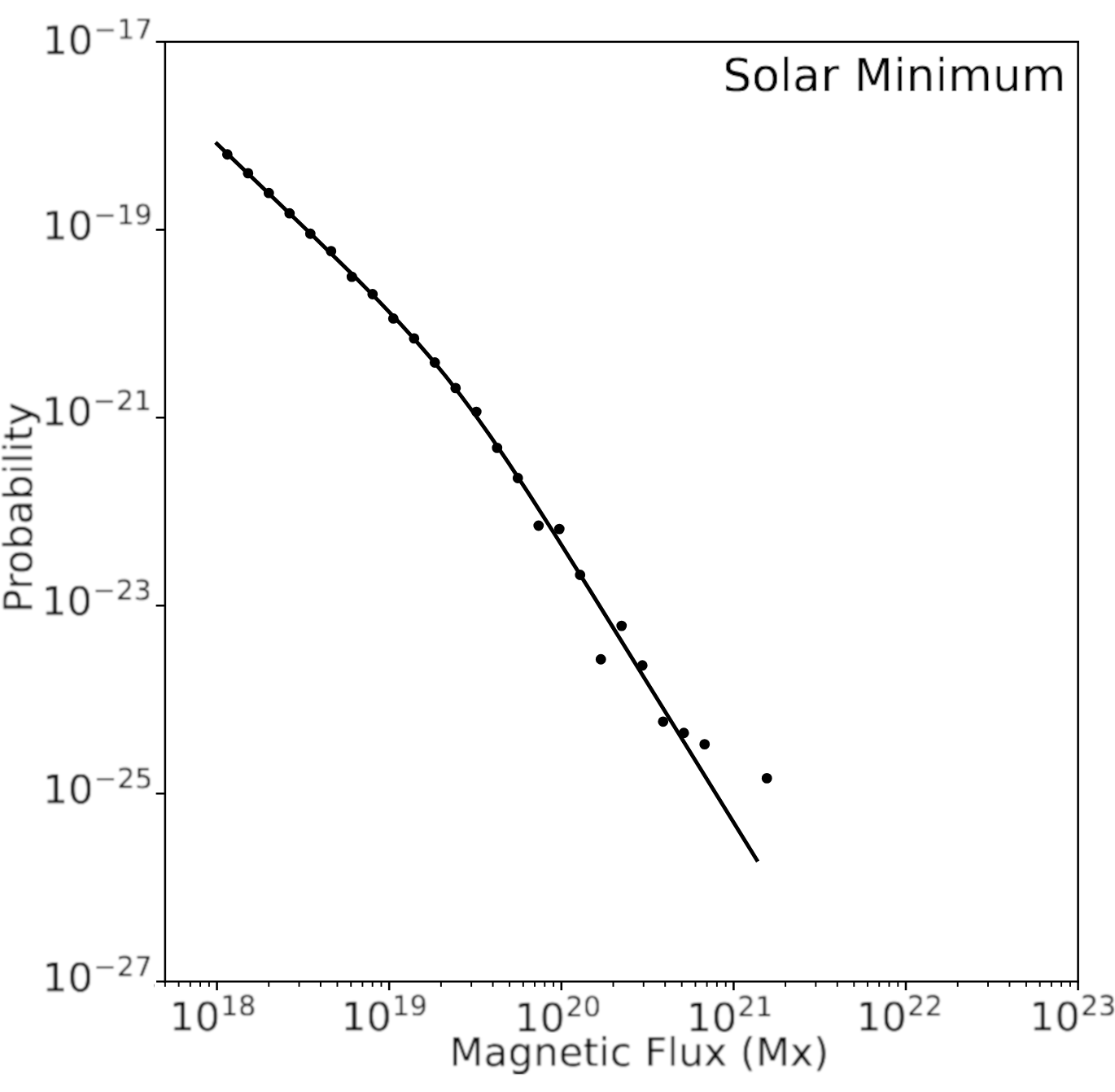}
    \end{center}  
    \setlength{\unitlength}{1cm}
    \begin{picture}(10,0.0)
    %Labels of figure location
    \put(0.65,6.53){{(a)}}
    \put(6.35,6.53){{(b)}}
    \put(0.65,0.9){{(c)}}
    \put(6.35,0.9){{(d)}}
    \end{picture}
	\caption{Histograms of the same data as in Figure \ref{fig:single_power_hists} but with a smooth double power law model fitted to the data rather than single power law model.}
	\label{fig:smooth_dpl_hists}
\end{figure}
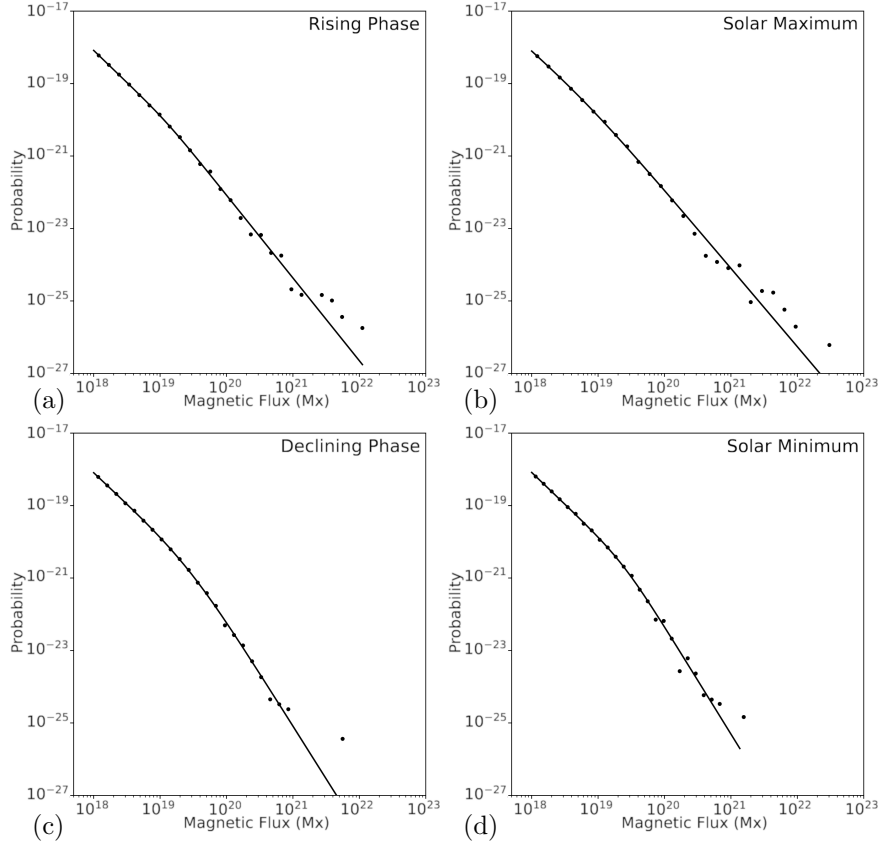

Therefore we choose the smooth double power law as the best representation of the true distribution of magnetic flux features over the course of a full solar cycle. 
Figure \ref{fig:smooth_dpl_hists} shows the histograms of magnetic flux distribution representing the same four phases of the solar cycle as in Figure \ref{fig:single_power_hists} but with a smooth double power law model overplotted on the histogram rather than a single power law model. The smooth double power law's flexibility allows it to capture the behaviour at the high flux tail to a much higher accuracy than the single power law.
We remark that this result is largely in accordance with the ``two-segment'' power law distribution 
put forward by \citet{song:etal24}, although their distribution seems to correspond to what we call
sharp double power law in this paper.

\section{Double power law: variation over one solar cycle}
    \label{s:Discussion}

\begin{figure}
   \begin{center}
     \includegraphics[width=\textwidth]{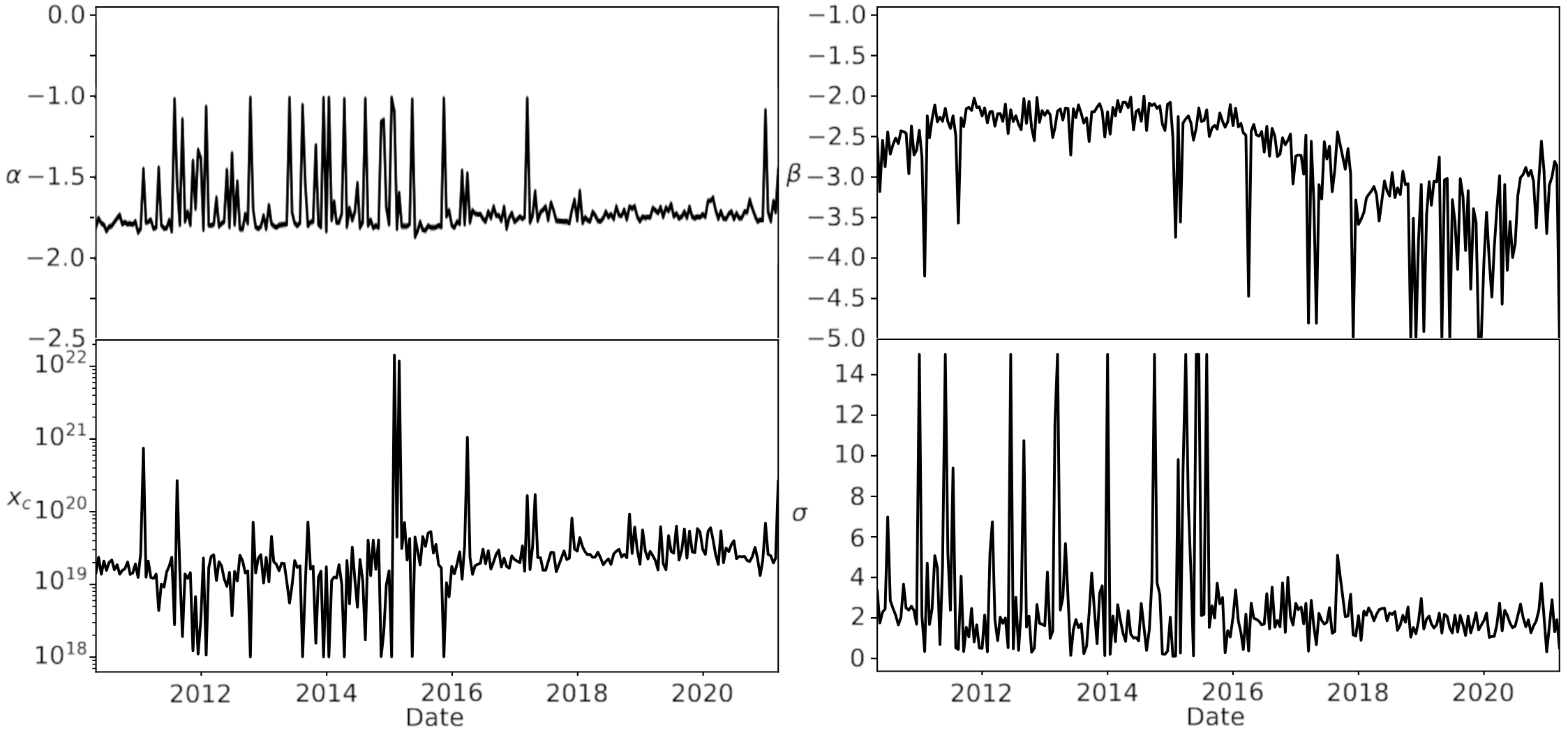} 
   \end{center}
    \setlength{\unitlength}{1cm}
    \begin{picture}(10,0.0)
    %Labels of figure location
    \put(5.58,6.15){\small{(a)}}
    \put(11.65,6.15){\small{(b)}}
    \put(5.58,3.57){\small{(c)}}
    \put(11.65,3.57){\small{(d)}}
    \end{picture}
    \caption{Temporal variation of the parameters of the smooth double power 
    law model over the solar cycle. Panel (a) shows $\alpha$, panel (b) shows $\beta$, 
    panel (c) shows $x_c$, and panel (d) shows $\sigma$.}
	\label{fig:cycle_params}
\end{figure}

A more in depth understanding of the double power law model can be obtained by investigating the temporal variation of the slope parameters of the separate power laws. Figure \ref{fig:cycle_params}  shows how the fitted parameters of the smooth double power law model vary over %time. 
one solar cycle.
%
% TN 26/10/2024: discussion of the other two
% parameters added from CNN PhD thesis.
%
% \textcolor{red}{(Just thinking out loud ... this 
% figure shows variation of all four parameters, 
% but we only talk about the power law slope parameters.
% Should we change the figure to reflect this, 
% or talk about the other parameters also?)}

Panel (a) shows the temporal variation of $\alpha$, the slope of the left hand power law. Typically, the value of $\alpha$ lies between $-1.7$ and $-1.8$ but there are plenty of upward fluctuations in the curve. There is much more consistency in the value at times of 
%less 
reduced
solar activity (from 2017 onwards reaching solar minimum in late 2019). Although
the fluctuations might simply be a reflection of statistical variations in the date,
one other possible explanation could be that the decay of large scale active regions does contribute to
the small scale flux distribution over shorter time scales. This would explain why these fluctuations seem
to occur mainly during enhanced solar activity.

Panel (b) shows the temporal variation of $\beta$, the slope of the right hand power law. There 
%seems to be some sort of 
is a
noticeable trend: the slope (magnitude of $\beta$) becomes steeper at solar minimum and 
%shallows out 
%shallower
flatter
at solar maximum. Again, there are a number of outliers in the graph. Interestingly, the spikes all point towards more negative values, whereas the spikes in the $\alpha$ curve all point towards less negative values. A value of $-5$ was set as the lower limit in the maximum likelihood estimation optimisation algorithm, so it is possible that the true value of 
$\beta$ in some magnetograms is less than $-5$. However, due to numerical limitation we have to compromise slightly on accuracy. For the most part, the value of $\beta$ is greater than $-5$. The upper envelope of
the curve seems to vary roughly between $-3.0$ (low activity) and $-2.0$ (high activity)

The temporal variation of the `kink location' ($x_c$) is shown in panel (c). Once again, the data is spiky but typically the value lies between $10^{19}$ and $10^{20}$ Mx. The values seem to be 
slightly
higher at solar minimum than at solar maximum 
but whether this is a real feature of the distribution or %just 
the result of statistical 
fluctuations is unclear. This question could be investigated
further in the future.

Finally, in panel (d) we see the temporal variation of the smoothing parameter ($\sigma$). A larger value of $\sigma$ means the transition from slope $\alpha$ to slope $\beta$ is sharper. Low values of $\sigma$ means that the transition 
%takes longer. 
takes place over a wider range.
Typically, the value lies between 2 and 4, however there are a number of spikes in either direction (both toward a value of 0 and toward a value of 15). A value of 15 was set as an upper limit for the maximum likelihood estimation optimisation algorithm. The spikes toward the highest value of $\sigma$ could be even higher, but we argue that a value of 15 is already sharp enough to convey the required information. In addition, computational limitations, again, mean we have to make a compromise on accuracy in these situations.

We see that there is much more variation in the 
$\beta$-parameter than the $\alpha$-parameter. This tells us that the slope of the small-scale flux 
distribution
is much more stable over a solar cycle than 
that of the large-scale flux distribution. 
The right hand slope ($\beta$) varies with 
changing solar activity; 
typically having a shallower slope at solar maximum and a steeper drop-off at solar minimum. 
This is to be expected because
the right hand power law slope ($\beta$) is 
affected by the presence of large-scale active regions.

The fact that the power law index of the small-scale flux distribution 
%is consistent 
appears to remain unchanged on average
regardless of the level of solar activity
%. This may 
could
be interpreted as an indication that the long-term
behaviour of the
small-scale flux distribution is not 
%the result of 
dominated by the
fragmentation of larger features, but largely due to
%some sort of 
a mechanism 
%leading to the generation of 
which perpetually generates
new, small-scale features throughout the entire solar cycle. 
% This may take the form of a dynamo action which is distributed throughout the entire convection zone, with large-scale features able to emerge at solar maximum due to the amplification of toroidal magnetic field stored in the tachocline.  

% We provide here a possible interpretation of the 
% observed magnetic flux distribution but are hesitant 
% to make definitive statements about the true nature of
% the mechanisms which lead to such a distribution. 
% We hope that this study may provide a statistical
% background upon which further theoretical progress 
% can be achieved.

%
% TN: 27/02/26 - comparison with Song et al. (2025) added.
%
We note that our results are qualitatively consistent with those of \citet{song:etal24}. 
Although we did not assume from the outset 
that a double power law distribution is the correct fit for the data, we found using
statistical goodness of fit tests that over a full solar cycle 
a smooth double power law is the best fitting distribution
function amongst those we tested. This very similar to the ``two-segment'' power law distribution fitted by 
\citet{song:etal24}, although this would correspond to our sharp double power law and not to the smooth
double power law we found to be the best fit over the full solar cycle. Quantitatively, we found
that our power law index for the small scale flux part ($\alpha$) is roughly between $-1.7$ and $-1.8$ with
more upward fluctuations during solar maximum, whereas \citet{song:etal24} quote a slightly larger average value of
$-1.64$. We note that it is possible that when averaging over the whole cycle we would arrive
at a higher value due to the upward fluctuations we find. However, given the nature of the variations
of $\alpha$ as shown in panel (a) of Figure \ref{fig:cycle_params} we do not believe that an average value
would necessarily be meaningful. Our value of the power law index for the large scale flux part ($\beta$)
is roughly in the same range, but at some times considerably lower than the corresponding 
values given by \citet{song:etal24}, which
vary between $-1.92$ at solar maximum und $-2.27$ at solar minimum. 
We would suggest that this difference could be due to the difference 
between the sharp double power law (``two-segment'' power law) used by \citet{song:etal24} and
the smooth double power law which we use. This is also reflected by the fact that \citet{song:etal24}
fix the transition point between the two segments of their double power law to 
$5.5 
%\cdot 
%TN revised: 15/06/2026
\times
10^{18}$Mx, whereas
we have a transition region whose position ($x_c$)and width ($\sigma$)is part of the fitting process. 
Despite these differences which we consider to be relatively minor, we find it very encouraging
that our findings using a statistical goodness of fit approach seem to give results that are
largely consistent with the findings by \citet{song:etal24}.

% The conclusion section 

\section{Summary and Conclusions}
    \label{s:Conclusions}

    % {\bf Need to rewrite/edit the text below}
    
    % In our study, we analysed 
    In this paper we have presented the analysis of 
    260 full disk line-of-sight HMI magnetograms over an 11 year period from May 2010 to March 2021. We used the modified clumping algorithm to identify individual magnetic flux features and generated size distributions
    of these features.
    We then used the maximum likelihood method to obtain fits to these 
    observational size distributions by nine different probablity density functions: 
    (single) power law PDF, 
    expontial PDF, 
    lognormal PDF, 
    Weibull PDF, 
    truncated Weibull PDF, 
    sharp double power law PDF,
    smooth double power law PDF,
    Weibull-lognormal PDF, and
    truncated Weibull-lognormal PDF.

% Double power law paragraph, inclding different criteria

Using a variety of statistical goodness-of-fit criteria we established that over a full solar cycle the smooth
double power law seems to represent the best fit to the observations. We analysed how the parameters of the
smooth double power law change over the solar cycle. On the basis of this analysis we found that the
power law index for the small flux features ($\alpha$) shows no long-term solar cycle variation, whereas the
power law index for the large flux features ($\beta$) shows a variation which is in line with the decrease
of large scale active region flux during solar minimum. The two other parameters of the smooth double 
power law PDF are the location of the transiation between the two power laws ($x_c$) and the 
width of the transition ($\sigma$). On the basis of our analysis we could not identify any significant long-term
variation of these two parameters over the full solar cycle.
    
The smooth double power law is an example of a bi-modal 
distribution. Bi-modal distributions are often taken
to be an indication that the two different parts of
the distribution could be generated by different
physical processes. If interpreted in such a way our result
could be viewed as being supportive of 
processes keeping the small-scale flux distribution largely time-invariant over a solar cycle, for example small-scale dynamo processes as discussed in \citet{rempel:etal23}, while the solar cycle variation
of the large-scale flux distribution is generated by a large-scale dynamo. However, one has to be cautious not
to overinterpret our results for a number of reasons. Firstly, the analysis presented in this paper is based on a 
single solar cycle only. It will need to be confirmed over other solar cycles, which clearly is a long-term project.
Secondly, our results do not exclude the possibility that a single physical mechanism could be 
responsible for both the relative temporal stability of the flux distribution at small scales while also
generating the variation of the large-scale distribution over the solar cycle. A solid theoretical understanding 
of the various physical processes is necessary to be able to distinguish between the different options. 
Finally, we note that the power indices as well as the other parameters of the flux distribution 
are subject to statistical fluctuations. While the overall long-term trend in the (large-scale) power law index $\beta$
is relatively obvious, we cannot exclude the possibilty that the statistical fluctuations could prevent 
%as 
%TN revised: 15/06/2026
us
from
detecting much smaller systematic variations in $\alpha$, for example. In this context, we mention the
work by \citet{korpi-lagg:etal22} who found no signficant solar cycle dependence of the 
quiet sun internetwork magnetic field 
fluctuations when using $1^\circ$ data patches, but clear solar cycle dependence of network and internetwork magnetic field 
fluctuations when using larger patches ($15^\circ$). Although the paper by \citet{korpi-lagg:etal22} studies
a different quantity than we did in the current paper, it shows that the question of solar cycle dependence 
of small scale magnetism is a subtle one.
Clearly, further studies of this important problem 
are warranted, both on the observational/data analysis side and on the theoretical side.

%%%%%%%%%%%%%%%%%%%%%%%%%%%%%%%%%%%%%%%%%%%%%%%%%%%%%%%%%%%%%%%%%%%%%%%%%%%
%% Appendix
%
\appendix

\section{Distribution Functions}
    \label{a:distributions}
    
    For each of the different
    distribution 
    functions
    (excluding the double power law distribution) investigated in Section \ref{s:Results}, we provide, here in Sections \ref{a:dists_sgl_power} to \ref{a:dists_log_normal}, details of the probability density function (PDF, $f(x; \theta_i)$), cumulative distribution function (CDF, $F(x; \theta_i)$), and an outline of how the maximum likelihood estimates (MLEs, $\hat{\theta}_i$) of the parameters ($\theta_i$ for $i = 1,2,\dots,m$; where $m$ is the number of parameters) are determined. Section \ref{a:dists_dbl_power} provides a more in-depth exploration of both forms of the double power law distribution; providing details of how the distribution is constructed and how the parameters may be estimated. It is important to note that the PDFs need to be multiplied by the total number of observed features in order to obtain the true frequency of fluxes.
    
    \subsection{Single Power Law}
        \label{a:dists_sgl_power}
        
        The form of the normalised PDF defined over $x \in [x_0, \infty)$ for the single power law distribution is:
        \begin{equation}
            f(x; \alpha) = \frac{\alpha - 1}{x_0}\left(\frac{x}{x_0}\right)^{-\alpha},
        \end{equation} provided $\alpha > 1$. The CDF of the single power law is defined as:
        \begin{equation}
            F(x; \alpha) = 1 - \left(\frac{x}{x_0}\right)^{1-\alpha}.
        \end{equation} The log-likelihood function for the single power law PDF is
        \begin{equation} \label{eq:power_log_lik}
            \mathcal{L}(\alpha) = \sum_{i=1}^N \left[\ln{\left(\alpha - 1\right)} - \ln{x_0} - \alpha\ln{\left(\frac{x_i}{x_0}\right)}\right],
        \end{equation} where N is the total number of data points, $x_i$. Setting the derivative of Equation (\ref{eq:power_log_lik}) with respect to $\alpha$ equal to zero and rearranging leads to the following formula for the MLE of $\alpha$:
        \begin{equation}
            \hat{\alpha} = 1 + \frac{N}{\sum_{i=1}^N \ln{\left(\frac{x_i}{x_0}\right)}}.
        \end{equation}
        
    \subsection{Weibull}
        \label{a:dists_weibull}
        
        The form of the normalised PDF defined over $x \in [x_0, \infty)$ for the Weibull distribution is:
        \begin{equation}
            f(x; \beta, \gamma) = \frac{\gamma}{\beta}\left(\frac{x - x_0}{\beta}\right)^{\gamma - 1}\exp{\left(-\left(\frac{x - x_0}{\beta}\right)^\gamma\right)},
        \end{equation} provided $\beta, \gamma > 0$ The CDF of the Weibull distribution is defined as:
        \begin{equation}
            F(x; \beta, \gamma) = 1 - \exp{\left(-\left(\frac{x - x_0}{\beta}\right)^\gamma\right)}.
        \end{equation} The log-likelihood function for the Weibull PDF is:
        \begin{equation} \label{eq:weibull_log_lik}
            \mathcal{L}(\beta, \gamma) = \sum_{i=1}^N \left[\ln{\left(\frac{\gamma}{\beta}\right)} + \left(\gamma - 1\right) \ln{\left(\frac{x_i - x_0}{\beta}\right)} - \left(\frac{x_i - x_0}{\beta}\right)^\gamma\right],
        \end{equation} where N is the total number of data points, $x_i$. Setting the derivatives of Equation (\ref{eq:weibull_log_lik}) with respect to both $\beta$ and $\gamma$ equal to zero and rearranging leads to the following formulae:
        \begin{equation} \label{eq:weibull_beta}
            \beta = \left(\frac{1}{N}\sum_{i=1}^N\left(x_i - x_0\right)^\gamma\right)^\frac{1}{\gamma},
        \end{equation}
        \begin{equation} \label{eq:weibull_gamma}
            \frac{1}{\gamma} + \frac{1}{N}\sum_{i=1}^N\ln{\left(x_i - x_0\right)} - \frac{\sum_{i=1}^N\left[\left(x_i - x_0\right)^\gamma\ln{\left(x_i - x_0\right)}\right]}{\sum_{i=1}^N\left(x_i - x_0\right)^\gamma} = 0.
        \end{equation} To obtain the MLE estimates for $\beta$ and $\gamma$ we first solve Equation (\ref{eq:weibull_gamma}) numerically to estimate $\hat{\gamma}$ and then determine $\hat{\beta}$ by substituting $\hat{\gamma}$ into Equation (\ref{eq:weibull_beta}).
        
    \subsection{Exponential}
        \label{a:dists_expon}
        
        The form of the normalised PDF defined over $[x_0, \infty)$ for the exponential distribution is:
        \begin{equation}
            f(x; \lambda) = \lambda e^{-\lambda\left(x - x_0\right)},
        \end{equation} provided $\lambda > 0$. The CDF of the exponential distribution is defined as:
        \begin{equation}
            F(x; \lambda) = 1 - e^{-\lambda\left(x - x_0\right)}.
        \end{equation} The log-likelihood function for the exponential PDF is:
        \begin{equation} \label{eq:expon_log_lik}
            \mathcal{L}(\lambda) = N\ln{\lambda} - \lambda\sum_{i = 1}^N\left(x_i - x_0\right),
        \end{equation} where N is the total number of data points, $x_i$. Setting the derivative of Equation (\ref{eq:expon_log_lik}) with respect to $\lambda$ equal to zero and rearranging leads to the following formula for the MLE of $\lambda$:
        \begin{equation}
            \hat{\lambda} = \frac{N}{\sum_{i=1}^N\left(x_i - x_0\right)}.
        \end{equation}
        
    \subsection{Log-normal}
        \label{a:dists_log_normal}

        The form of the normalised PDF defined over $[x_0, \infty)$ for the log-normal distribution is:
        \begin{equation}
            f(x; \mu, \sigma) = \frac{1}{x\sigma\sqrt{2\pi}}\exp{\left(-\frac{\left(\ln{x} - \mu\right)^2}{2\sigma^2}\right)},
        \end{equation} provided $\sigma > 0$. The CDF of the log-normal distribution is defined as:
        \begin{equation}
            F(x; \mu, \sigma) = \frac{1}{2}\left[1 - \mathrm{erf}\left(\frac{\ln{x} - \mu}{\sigma\sqrt{2}}\right)\right],
        \end{equation} where $\mathrm{erf}$ is the error function. The log-likelihood function for the log-normal PDF is:
        \begin{equation} \label{eq:lognormal_log_lik}
            \mathcal{L}(\mu, \sigma) = -N\ln{\sigma} - N\ln{\sqrt{2\pi}} - \sum_{i=1}^N\ln{x_i} - \frac{\sum_{i=1}^N\left(\ln{x_i} - \mu\right)^2}{2\sigma^2},
        \end{equation} where N is the total number of data points, $x_i$. Taking the derivative of Equation (\ref{eq:lognormal_log_lik}) with respect to both $\mu$ and $\sigma$ leads to the following formula for the MLEs of $\mu$ and $\sigma$:
        \begin{equation}
            \hat{\mu} = \frac{\sum_{i=1}^N \ln{x_i}}{N}
        \end{equation}
        \begin{equation}
            \hat{\sigma} = \sqrt{ \frac{\sum_{i=1}^N \left(\ln{x_i} - \hat{\mu}\right)^2}{N} }.
        \end{equation}
        
    \subsection{Double Power Law}
        \label{a:dists_dbl_power}
        
\begin{figure}
            \centerline{
             \includegraphics[width=0.6\textwidth,clip=]{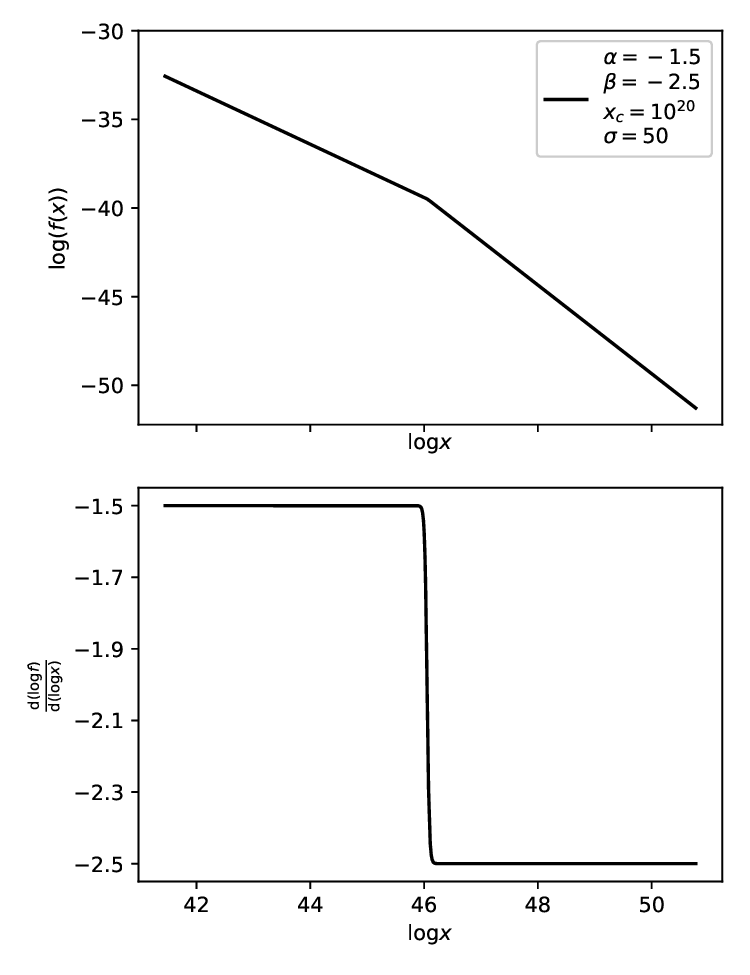}}
            \vspace{-0.75\textwidth}
                \centerline{\Large \bf
                \hspace{0.15\textwidth}  (a)
                \hfill}
                \vspace{0.35\textwidth}
                \centerline{\Large \bf 
                \hspace{0.15\textwidth} (b)
                \hfill}
                \vspace{0.3\textwidth}
            \caption{(a) A typical double power law will have the same structure as shown in the graph, constrained by four parameters; on a log-log scale the function will display two different straight line behaviours. The lines represent different power laws which transition from one to the other at some location, $x_c$, which we call the `kink' location. To the left of the kink we have a power law with index $\alpha$ and to the right we have a power law with index $\beta$. The rate at which the function transitions from one power law to the other is controlled by a fourth parameter, $\sigma$; larger values of $\sigma$ lead to a quicker transition and hence a sharper kink. (b) The derivative of the function (in log-log space) allows us to mathematically formulate the double power law function. We model the derivative as a hyperbolic tangent function shifted in the $y$-axis which is constrained by the four parameters; $\alpha$ and $\beta$ which control the amplitude and shift in $y$-axis, $x_c$ which controls the location, and $\sigma$ which controls the width of the function.}
            \label{fig:dbl_power_cartoon}
        \end{figure}

        The double power law distribution describes data which follow separate power laws in separate sections of the distribution's domain, i.e. two separate power laws with different indices, $\alpha$ and $\beta$, describing two separate sections of the data, respectively. The location at which the function transitions from one power law form to the other, known as the `kink' location, is denoted by the parameter $x_c$ and the rate at which the function transitions in controlled by a fourth parameter, $\sigma$. Determining the mathematical form of such a function is not straight-forward but is significantly easier if one considers the functional form of its derivative and integrate up to obtain the correct PDF. Figure \ref{fig:dbl_power_cartoon} shows a cartoon illustration of a typical double power law and its derivative. In log-log space we can model the derivative of the function, $\frac{\mathrm{d}(\log{(f(x))})}{\mathrm{d}(\log{x})}$, as a hyperbolic tangent function constrained by the four parameters as follows:
        
        \begin{equation} \label{eq:dbl_power_deriv}
            \frac{\mathrm{d}[\log{(f(x; \alpha, \beta, x_c, \sigma))}]}{\mathrm{d}\phi} = \frac{\alpha - \beta}{2}\tanh{\left(\frac{-\sigma\left(\phi - \phi_c\right)}{2}\right)} + \frac{\alpha + \beta}{2},
        \end{equation} 
        where $\phi = \log{x}$ and $\phi_c = \log{x_c}$. Integrating Equation \ref{eq:dbl_power_deriv} with respect to $\phi$ and substituting $x$ back in leads to the following function for the double power law PDF:
        \begin{equation} \label{eq:dbl_power_pdf}
            f(x; \alpha, \beta, x_c, \sigma) = K\frac{x_c^{\frac{\alpha + \beta}{2}}}{2^{\frac{\beta - \alpha}{\sigma}}}\left(\frac{x}{x_c}\right)^\alpha\left[1 + \left(\frac{x}{x_c}\right)^\sigma\right]^{\frac{\beta - \alpha}{\sigma}},
        \end{equation} 
        where $K$ is a normalisation constant.
        
        The normalisation of the function can be treated in two separate ways. The first method treats $\sigma$ as a fixed value which tends to $\infty$, causing a sharp kink in the distribution at $x_c$ (Sharp Double Power Law, Sharp-DPL). The second method keeps $\sigma$ as a variable parameter allowing for a smoother transition between power laws (Smooth Double Power Law, Smooth-DPL). The following sections detail how each separate method is treated.
        
        \subsubsection{Sharp Double Power Law}
            The Sharp Double Power Law (Sharp-DPL) is formulated by allowing $\sigma \rightarrow \infty$. The function in Equation \ref{eq:dbl_power_pdf} reduces to the following piece-wise function:
            \begin{equation} \label{eq:Sharp-DPL non-normalised}
                f(x; \alpha, \beta, x_c) = K x_c^{\frac{\alpha+\beta}{2}}\left\{
                    \begin{array}{ll}
                        \left(\frac{x}{x_c}\right)^\alpha, & \quad x \leq x_c \\
                        \left(\frac{x}{x_c}\right)^\beta, & \quad x > x_c. \\
                    \end{array}
                \right.
            \end{equation} Equation \ref{eq:Sharp-DPL non-normalised} can be normalised analytically resulting in the following PDF:
            \begin{equation} \label{eq:Sharp-DPL normalised}
                f(x; \alpha, \beta, x_c) = \frac{\left(\alpha+1\right)\left(\beta + 1\right)}{x_c\left[\beta-\alpha-\left(\beta+1\right)\left(\frac{x_0}{x_c}\right)^{\alpha+1}\right]} \left\{
                    \begin{array}{ll}
                        \left(\frac{x}{x_c}\right)^\alpha, & \quad x \leq x_c \\
                        \left(\frac{x}{x_c}\right)^\beta, & \quad x > x_c, \\
                    \end{array}
                \right.
            \end{equation} provided $\alpha, \beta < -1$. It is not feasible to determine the maximum likelihood estimates of the three parameters ($\alpha$, $\beta$ and $x_c$) analytically so instead one has to perform a numerical optimisation of the log-likelihood function of the PDF. The CDF of the function is as follows:
            \begin{equation} \label{eq:Sharp-DPL CDF}
                F(x; \alpha, \beta, x_c) = K' \left\{
                    \begin{array}{ll}
                        \left[\left(\frac{x}{x_c}\right)^{\alpha+1} - \left(\frac{x_0}{x_c}\right)^{\alpha+1}\right] / \left(\alpha+1\right), & \quad x \leq x_c \\
                        \frac{\left(\beta+1\right)\left[1 - \left(\frac{x_0}{x_c}\right)^{\alpha+1}\right] + \left(\alpha+1\right) \left[\left(\frac{x}{x_c}\right)^{\beta+1} - 1\right]}{\left(\alpha+1\right)\left(\beta+1\right)} , & \quad x > x_c, \\
                    \end{array}
                \right.
            \end{equation} where $K' = \frac{\left(\alpha+1\right)\left(\beta + 1\right)}{\beta-\alpha-\left(\beta+1\right)\left(\frac{x_0}{x_c}\right)^{\alpha+1}} $
            
        \subsubsection{Smooth Double Power Law}
            The Smooth Double Power Law treats $\sigma$ as a variable parameter which introduces a smooth transition between separate power laws. Direct integration of Equation \ref{eq:dbl_power_pdf} introduces hypergeometric functions which makes the normalisation of the function difficult to approach. An alternative approach is to perform a numerical normalisation of the function by re-writing the PDF as follows:
            \begin{equation}
                f(x; \alpha, \beta, x_c, \sigma) = \frac{I\left(x; \alpha, \beta, x_c, \sigma\right)}{\int_{x_0}^\infty I\left(x; \alpha, \beta, x_c, \sigma\right)\mathrm{d}x},
            \end{equation} where $$ I\left(x; \alpha, \beta, x_c, \sigma\right) = \left(\frac{x}{x_c}\right)^\alpha\left[1+\left(\frac{x}{x_c}\right)^\sigma\right]^{\frac{\beta-\alpha}{\sigma}}.$$
            One can then determine the maximum likelihood estimates of all four parameters ($\alpha$, $\beta$, $x_c$ and $\sigma$) using a combination of numerical optimisation of the log-likelihood of the above function and numerical integration of the stated integrand. The CDF is also obtained through numerical integration.

%%%%%%%%%%%%%%%%%%%%%%%%%%%%%%%%%%%%%%%%%%%%%%%%%%%%%%%%%%%%%%%%%%%%%%%%%%%
%% Acknowledgements
%
\begin{acks}
CNN acknowledges support by the UK’s Science 
and Technology Facilities Council (STFC) 
Doctoral Training 
Centre Grant ST/P006809/1 (ScotDIST). 
TN and CEP both acknowledge 
support by the STFC Consolidated Grants 
ST/S000402/1 and ST/W001195/1.
\end{acks}

%% Available additional data environments:
%% required: authorcontribution, fundinginformation, dataavailability
%% optional: materialsavailability, codeavailability
 \begin{authorcontribution}
 CNN undertook the statistical analysis of the data. All authors contributed to 
 defining the scientific questions, selecting the methods, the interpretation of the results and the writing of the paper.
 \end{authorcontribution}
 \begin{fundinginformation}
This work was supported by the UK's Science and Technology
Facilities Council (STFC) via Doctoral Training 
Centre Grant ST/P006809/1 (ScotDIST), Consolidated Grants ST/S000402/1 and ST/W001195/1.
 \end{fundinginformation}
 \begin{dataavailability}
The data used in this paper are publicly available at 
\linebreak
http://jsoc.stanford.edu/.
 \end{dataavailability}
 \begin{ethics}
 \begin{conflict}
The authors declare no competing interests.
 \end{conflict}
 \end{ethics}

%%% %%%%%%%%%%%%%%%%%%%%%%%%%%%%%%%%%%%%%%%%%%%%%%%%%%%%%%%%%%%
%% Bibliography
%
% Using BibTeX
%
\bibliographystyle{spr-mp-sola}
\bibliography{biblio}

@article{baumann2005size,
  title={On the size distribution of sunspot groups in the {G}reenwich sunspot record 1874--1976},
  author={Baumann, Ingo and Solanki, SK},
  journal={\aap},
  volume={443},
  pages={1061--1066},
  year={2005},
  publisher={EDP Sciences}
}

@ARTICLE{bellotrubio2019livrev,
       author = {{Bellot Rubio}, Luis and {Orozco Su{\'a}rez}, David},
        title = "{Quiet Sun magnetic fields: an observational view}",
      journal = {\lrsp},
         year = 2019,
        month = feb,
       volume = {16},
          eid = {1},
        pages = {1},
          doi = {10.1007/s41116-018-0017-1},
       adsurl = {https://ui.adsabs.harvard.edu/abs/2019LRSP...16....1B}
}

@article{bogdan1988distribution,
  title={Distribution of sunspot umbral areas 1917-1982},
  author={Bogdan, T. J. and Gilman, Peter A and Lerche, I and Howard, Robert},
  journal={\apj},
  volume={327},
  pages={451--456},
  year={1988}
}

@article{brandenburg2012current,
  title={Current status of turbulent dynamo theory},
  author={Brandenburg, Axel and Sokoloff, Dmitry and Subramanian, Kandaswamy},
  journal={\ssr},
  volume={169},
  pages={123--157},
  year={2012},
  publisher={Springer}
}

@ARTICLE{brun2017livrev,
       author = {{Brun}, Allan Sacha and {Browning}, Matthew K.},
        title = "{Magnetism, dynamo action and the solar-stellar connection}",
      journal = {\lrsp},
         year = 2017,
        month = sep,
       volume = {14},
          eid = {4},
        pages = {4},
          doi = {10.1007/s41116-017-0007-8},
       adsurl = {https://ui.adsabs.harvard.edu/abs/2017LRSP...14....4B}
}

@book{chakravarti:etal1967,
    author = {{Chakravarti}, I. M. and {Laha}, R. G. and {Roy}, J.}, 
    title = {Handbook of Methods of Applied Statistics}, 
    volume = "Volume I",
    publisher = {John Wiley and Sons},
    address = {New York},
    year = {1967}
}

@article{charbonneau2014annual,
  author = {Charbonneau, Paul},
  title = {Solar Dynamo Theory},
  journal = {\araa},
  volume = {52},
  pages = {251-290},
  year = {2014},
  doi = {10.1146/annurev-astro-081913-040012}
}

@ARTICLE{charboneau2020livrev,
       author = {{Charbonneau}, Paul},
        title = "{Dynamo models of the solar cycle}",
      journal = {\lrsp},
         year = 2020,
        month = jun,
       volume = {17},
          eid = {4},
        pages = {4},
          doi = {10.1007/s41116-020-00025-6},
       adsurl = {https://ui.adsabs.harvard.edu/abs/2020LRSP...17....4C}
}

@ARTICLE{cho2015,
       author = {{Cho}, I. -H. and {Cho}, K. -S. and {Bong}, S. -C. and {Lim}, E. -K. and {Kim}, R. -S. and {Choi}, S. and {Kim}, Y. -H. and {Yurchyshyn}, V.},
        title = "{Statistical Comparison Between Pores and Sunspots by Using SDO/HMI}",
      journal = {\apj},
         year = 2015,
        month = sep,
       volume = {811},
          eid = {49},
        pages = {49},
          doi = {10.1088/0004-637X/811/1/49},
       adsurl = {https://ui.adsabs.harvard.edu/abs/2015ApJ...811...49C}
}

@article{clauset2009power,
  title={Power-law distributions in empirical data},
  author={Clauset, Aaron and Shalizi, Cosma Rohilla and Newman, Mark E J},
  journal={SIAM review},
  volume={51},
  pages={661--703},
  year={2009},
  publisher={SIAM}
}

@article{deforest2007solar,
  title={Solar magnetic tracking. {I}. {S}oftware comparison and recommended practices},
  author={DeForest, CE and Hagenaar, HJ and Lamb, DA and Parnell, CE and Welsch, BT},
  journal={\apj},
  volume={666},
  pages={576},
  year={2007},
  publisher={IOP Publishing}
}

@book{gibbons2020nonparametric,
  title={{Nonparametric Statistical Inference}},
  author={Gibbons, Jean Dickinson and Chakraborti, Subhabrata},
  year={2020},
  publisher={CRC press},
  address = {New York}
}

@article{hagenaar1999dispersal,
  title={Dispersal of magnetic flux in the quiet solar photosphere},
  author={Hagenaar, H J and Schrijver, C J and Shine, R A and others},
  journal={\apj},
  volume={511},
  pages={932},
  year={1999},
  publisher={IOP Publishing}
}

@article{harvey1993properties,
  title={Properties and emergence patterns of bipolar active regions},
  author={Harvey, Karen L and Zwaan, Cornelis},
  journal={\solphys},
  volume={148},
  pages={85--118},
  year={1993},
  publisher={Springer}
}

@article{Hathaway_2015,
   title={The Solar Cycle},
   volume={12},
   ISSN={1614-4961},
   url={http://dx.doi.org/10.1007/lrsp-2015-4},
   DOI={10.1007/lrsp-2015-4},
   journal={\lrsp},
   publisher={Springer Science and Business Media LLC},
   author={Hathaway, David H.},
   year={2015},
   month={Sep}
}

@ARTICLE{houtgast1948,
       author = {{Houtgast}, J. and {van Sluiters}, A.},
        title = "{Statistical investigations concerning the magnetic fields of sunspots I}",
      journal = {\bain},
         year = 1948,
        month = mar,
       volume = {10},
        pages = {325},
       adsurl = {https://ui.adsabs.harvard.edu/abs/1948BAN....10..325H}
}

@ARTICLE{javaherian2017,
       author = {{Javaherian}, M. and {Safari}, H. and {Dadashi}, N. and {Aschwanden}, M.~J.},
        title = "{Statistical Properties of Photospheric Magnetic Elements Observed by the Helioseismic and Magnetic Imager onboard the Solar Dynamics Observatory}",
      journal = {\solphys},
         year = 2017,
        month = nov,
       volume = {292},
          eid = {164},
        pages = {164},
          doi = {10.1007/s11207-017-1189-x},
       adsurl = {https://ui.adsabs.harvard.edu/abs/2017SoPh..292..164J}
}

@article{jiang2011solar,
  title={The solar magnetic field since 1700 - {II}. {P}hysical reconstruction of total, polar and open flux},
  author={Jiang, Jie and Cameron, Robert H and Schmitt, Dieter and Sch{\"u}ssler, M},
  journal={\aap},
  volume={528},
  pages={A83},
  year={2011},
  publisher={EDP Sciences}
}

@ARTICLE{korpi-lagg:etal22,
       author = {{Korpi-Lagg}, M.~J. and {Korpi-Lagg}, A. and {Olspert}, N. and {Truong}, H. -L.},
        title = "{Solar-cycle variation of quiet-Sun magnetism and surface gravity oscillation mode}",
      journal = {\aap},
         year = 2022,
        month = sep,
       volume = {665},
          eid = {A141},
        pages = {A141},
          doi = {10.1051/0004-6361/202243979},
archivePrefix = {arXiv},
       eprint = {2205.04419},
 primaryClass = {astro-ph.SR},
       adsurl = {https://ui.adsabs.harvard.edu/abs/2022A&A...665A.141K}
}

@ARTICLE{kostyuchenko2017,
       author = {{Kostyuchenko}, I.~G.},
        title = "{Dynamic Characteristics of Area Variations of Small and Large Sunspots and Quasi-Biennial Oscillations in Solar Activity}",
      journal = {\gaa},
         year = 2017,
        month = dec,
       volume = {57},
        pages = {814-820},
          doi = {10.1134/S0016793217070118},
       adsurl = {https://ui.adsabs.harvard.edu/abs/2017Ge&Ae..57..814K}
}

@article{kuklin1980astronomical,
  title={On two populations of sunspot groups},
  author={Kuklin, G V},
  journal={Bulletin of the Astronomical Institutes of Czechoslovakia},
  volume={31},
  pages={224--232},
  year={1980}
}

@ARTICLE{kumar:etal25,
       author = {{Kumar}, Avneesh and {Kumar}, Nagendra and {Vats}, Hari Om},
        title = "{Statistical Analysis of Sunspot Area and Magnetic Flux on Solar Disc During 2011 {\textendash} 2023}",
      journal = {\solphys},
         year = 2025,
        month = sep,
       volume = {300},
          eid = {122},
        pages = {122},
          doi = {10.1007/s11207-025-02537-6},
       adsurl = {https://ui.adsabs.harvard.edu/abs/2025SoPh..300..122K}
}

@inproceedings{liu2016hmi,
  title={On {HMI}'s Mod-{L} Sequence: Test and Evaluation},
  author={Liu, Yang and Baldner, Charles and Bogart, R S and Bush, R and Couvidat, S and Duvall, Thomas L and Hoeksema, Jon Todd and Norton, Aimee Ann and Scherrer, Philip H and Schou, Jesper},
  booktitle={AAS/\solphys Division Abstracts\# 47},
  volume={47},
  year={2016}
}

@ARTICLE{mackay2012livrev,
       author = {{Mackay}, Duncan H. and {Yeates}, Anthony R.},
        title = "{The Sun's Global Photospheric and Coronal Magnetic Fields: Observations and Models}",
      journal = {\lrsp},
         year = 2012,
        month = nov,
       volume = {9},
          eid = {6},
        pages = {6},
          doi = {10.12942/lrsp-2012-6},
archivePrefix = {arXiv},
       eprint = {1211.6545},
 primaryClass = {astro-ph.SR},
       adsurl = {https://ui.adsabs.harvard.edu/abs/2012LRSP....9....6M}
}

@article{meunier2003statistical,
  title={Statistical properties of magnetic structures: Their dependence on scale and solar activity},
  author={Meunier, N},
  journal={\aap},
  volume={405},
  pages={1107--1120},
  year={2003},
  publisher={EDP Sciences}
}

@article{munoz2015small,
  title={Small-scale and global dynamos and the area and flux distributions of active regions, sunspot groups, and sunspots: A multi-database study},
  author={Mu{\~n}oz-Jaramillo, Andr{\'e}s and Senkpeil, Ryan R and Windmueller, John C and Amouzou, Ernest C and Longcope, Dana W and Tlatov, Andrey G and Nagovitsyn, Yury A and Pevtsov, Alexei A and Chapman, Gary A and Cookson, Angela M and others},
  journal={\apj},
  volume={800},
  pages={48},
  year={2015},
  publisher={IOP Publishing}
}

@article{nagovitsyn2012possible,
  title={On a possible explanation of the long-term decrease in sunspot field strength},
  author={Nagovitsyn, Yury A and Pevtsov, Alexei A and Livingston, William C},
  journal={\apjl},
  volume={758},
  pages={L20},
  year={2012},
  publisher={IOP Publishing}
}

@ARTICLE{nagovitsyn2017,
       author = {{Nagovitsyn}, Y.~A. and {Pevtsov}, A.~A. and {Osipova}, A.~A.},
        title = "{Long-term variations in sunspot magnetic field-area relation}",
      journal = {Astronomische Nachrichten},
         year = 2017,
        month = jan,
       volume = {338},
        pages = {26-34},
          doi = {10.1002/asna.201613035},
archivePrefix = {arXiv},
       eprint = {1608.01132},
 primaryClass = {astro-ph.SR},
       adsurl = {https://ui.adsabs.harvard.edu/abs/2017AN....338...26N}
}

@ARTICLE{nagovitsyn2021,
       author = {{Nagovitsyn}, Yury A. and {Pevtsov}, Alexei A.},
        title = "{Bi-lognormal Distribution of Sunspot Group Areas}",
      journal = {\apj},
         year = 2021,
        month = jan,
       volume = {906},
          eid = {27},
        pages = {27},
          doi = {10.3847/1538-4357/abc82d},
       adsurl = {https://ui.adsabs.harvard.edu/abs/2021ApJ...906...27N}
}

@article{newman2005power,
  title={Power laws, {Pareto} distributions and {Zipf's} law},
  author={Newman, Mark E J},
  journal={\contphys},
  volume={46},
  pages={323--351},
  year={2005},
  month = {October},
  doi ={10.1080/00107510500052444}
}

@ARTICLE{nicholson1933,
       author = {{Nicholson}, S.~B.},
        title = "{The Area of a Sun-Spot and the Intensity of Its Magnetic Field}",
      journal = {\pasp},
         year = 1933,
        month = feb,
       volume = {45},
        pages = {51-53},
          doi = {10.1086/124301},
       adsurl = {https://ui.adsabs.harvard.edu/abs/1933PASP...45...51N}
}

@ARTICLE{nikbaksh2019,
       author = {{Nikbakhsh}, S. and {Tanskanen}, E.~I. and {K{\"a}pyl{\"a}}, M.~J. and {Hackman}, T.},
        title = "{Differences in the solar cycle variability of simple and complex active regions during 1996-2018}",
      journal = {\aap},
         year = 2019,
        month = sep,
       volume = {629},
          eid = {A45},
        pages = {A45},
          doi = {10.1051/0004-6361/201935486},
archivePrefix = {arXiv},
       eprint = {1908.02226},
 primaryClass = {astro-ph.SR},
       adsurl = {https://ui.adsabs.harvard.edu/abs/2019A&A...629A..45N}
}

@article{parnell2002nature,
  title={Nature of the magnetic carpet--{I}. {D}istribution of magnetic fluxes},
  author={Parnell, C E},
  journal={\mnras},
  volume={335},
  pages={389--398},
  year={2002},
  publisher={The Royal Astronomical Society}
}

@article{parnell2009power,
  title={A power-law distribution of solar magnetic fields over more than five decades in flux},
  author={Parnell, C E and DeForest, C E and Hagenaar, H J and Johnston, B A and Lamb, D A and Welsch, B T},
  journal={\apj},
  volume={698},
  pages={75},
  year={2009},
  publisher={IOP Publishing}
}

@ARTICLE{petrie2013,
       author = {{Petrie}, G.~J.~D.},
        title = "{Solar Magnetic Activity Cycles, Coronal Potential Field Models and Eruption Rates}",
      journal = {\apj},
         year = 2013,
        month = may,
       volume = {768},
          eid = {162},
        pages = {162},
          doi = {10.1088/0004-637X/768/2/162},
archivePrefix = {arXiv},
       eprint = {1303.1218},
 primaryClass = {astro-ph.SR},
       adsurl = {https://ui.adsabs.harvard.edu/abs/2013ApJ...768..162P}
}

@ARTICLE{pevtsov2014,
       author = {{Pevtsov}, Alexei A. and {Bertello}, Luca and {Tlatov}, Andrey G. and {Kilcik}, Ali and {Nagovitsyn}, Yury A. and {Cliver}, Edward W.},
        title = "{Cyclic and Long-Term Variation of Sunspot Magnetic Fields}",
      journal = {\solphys},
         year = 2014,
        month = feb,
       volume = {289},
        pages = {593-602},
          doi = {10.1007/s11207-012-0220-5},
archivePrefix = {arXiv},
       eprint = {1301.5935},
 primaryClass = {astro-ph.SR},
       adsurl = {https://ui.adsabs.harvard.edu/abs/2014SoPh..289..593P}
}

@ARTICLE{pevtsov2021,
       author = {{Pevtsov}, Alexei A. and {Bertello}, Luca and {Nagovitsyn}, Yury A. and {Tlatov}, Andrey G. and {Pipin}, Valery V.},
        title = "{Long-term studies of photospheric magnetic fields on the Sun}",
      journal = {\jswsc},
         year = 2021,
        month = jan,
       volume = {11},
          eid = {4},
        pages = {4},
          doi = {10.1051/swsc/2020069},
       adsurl = {https://ui.adsabs.harvard.edu/abs/2021JSWSC..11....4P}
}

@ARTICLE{rempel:etal23,
       author = {{Rempel}, Matthias and {Bhatia}, Tanayveer and {Bellot Rubio}, Luis and {Korpi-Lagg}, Maarit J.},
        title = "{Small-Scale Dynamos: From Idealized Models to Solar and Stellar Applications}",
      journal = {\ssr},
         year = 2023,
        month = aug,
       volume = {219},
          eid = {36},
        pages = {36},
          doi = {10.1007/s11214-023-00981-z},
archivePrefix = {arXiv},
       eprint = {2305.02787},
 primaryClass = {astro-ph.SR},
       adsurl = {https://ui.adsabs.harvard.edu/abs/2023SSRv..219...36R}
}

@ARTICLE{sakurai:toriumi23,
       author = {{Sakurai}, Takashi and {Toriumi}, Shin},
        title = "{Probability Distribution Functions of Sunspot Magnetic Flux}",
      journal = {\apj},
         year = 2023,
        month = jan,
       volume = {943},
          eid = {10},
        pages = {10},
          doi = {10.3847/1538-4357/aca28a},
archivePrefix = {arXiv},
       eprint = {2211.13957},
 primaryClass = {astro-ph.SR},
       adsurl = {https://ui.adsabs.harvard.edu/abs/2023ApJ...943...10S}
}

@article{schad2010structural,
  title={Structural invariance of sunspot umbrae over the solar cycle: 1993--2004},
  author={Schad, T A and Penn, M J},
  journal={\solphys},
  volume={262},
  pages={19--33},
  year={2010},
  publisher={Springer}
}

@article{scholz2014maximum,
  title={Maximum likelihood estimation},
  author={Scholz, Fritz W},
  journal={Wiley StatsRef: Statistics Reference Online},
  year={2014},
  month = {September},
  doi = {10.1002/9781118445112.stat01663}
}

@article{schrijver1997sustaining,
  title={Sustaining the quiet photospheric network: The balance of flux emergence, fragmentation, merging, and cancellation},
  author={Schrijver, Carolus J and van Ballegooijen, Adriaan A and Hagenaar, Hermance J and Shine, Richard A and others},
  journal={\apj},
  volume={487},
  pages={424},
  year={1997},
  publisher={IOP Publishing}
}

@ARTICLE{shapoval2018,
       author = {{Shapoval}, A. and {Le Mou{\"e}l}, J. -L. and {Shnirman}, M. and {Courtillot}, V.},
        title = "{Observational evidence in favor of scale-free evolution of sunspot groups}",
      journal = {\aap},
         year = 2018,
        month = nov,
       volume = {618},
          eid = {A183},
        pages = {A183},
          doi = {10.1051/0004-6361/201832799},
       adsurl = {https://ui.adsabs.harvard.edu/abs/2018A&A...618A.183S}
}

@article{solanki2006solar,
  title={The solar magnetic field},
  author={Solanki, Sami K and Inhester, Bernd and Sch{\"u}ssler, Manfred},
  journal={\rpp},
  volume={69},
  pages={563},
  year={2006},
  publisher={IOP Publishing}
}

@ARTICLE{song:etal24,
       author = {{Song}, Anchuan and {Zhang}, Quanhao and {Wang}, Yuming and {Liu}, Rui and {Jiang}, Jie and {Li}, Xiaolei and {Liu}, Jiajia and {Lv}, Shaoyu and {Zheng}, Ruobing},
        title = "{Solar cycle variation in the properties of photospheric magnetic concentrations}",
      journal = {\aap},
         year = 2024,
        month = feb,
       volume = {682},
          eid = {A87},
        pages = {A87},
          doi = {10.1051/0004-6361/202346898},
       adsurl = {https://ui.adsabs.harvard.edu/abs/2024A&A...682A..87S}
}

@PHDTHESIS{strous1994dynamics,
       author = {{Strous}, L.~H.},
        title = {Dynamics in Solar Active Regions: Patterns in Magnetic-Flux Emergence},
       school = {Utrecht University},
         year = {1994},
        month = {Jun},
       adsurl = {https://ui.adsabs.harvard.edu/abs/1994PhDT.......347S}
}

@article{tang1984statistical,
  title={A statistical study of active regions 1967--1981},
  author={Tang, Frances and Howard, Robert and Adkins, John M},
  journal={\solphys},
  volume={91},
  pages={75--86},
  year={1984},
  publisher={Springer}
}

@article{thornton2011small,
  title={Small-scale flux emergence observed using {H}inode/{SOT}},
  author={Thornton, L M and Parnell, C E},
  journal={\solphys},
  volume={269},
  pages={13--40},
  year={2011},
  publisher={Springer}
}

@ARTICLE{tlatov2019,
       author = {{Tlatov}, Andrey and {Riehokainen}, Alexandr and {Tlatova}, Kseniya},
        title = "{The Characteristic Sizes of the Sunspots and Pores in Solar Cycle 24}",
      journal = {\solphys},
         year = 2019,
        month = apr,
       volume = {294},
          eid = {45},
        pages = {45},
          doi = {10.1007/s11207-019-1439-1},
       adsurl = {https://ui.adsabs.harvard.edu/abs/2019SoPh..294...45T}
}

@ARTICLE{wang:etal23,
       author = {{Wang}, Ruihui and {Jiang}, Jie and {Luo}, Yukun},
        title = "{Toward a Live Homogeneous Database of Solar Active Regions Based on SOHO/MDI and SDO/HMI Synoptic Magnetograms. I. Automatic Detection and Calibration}",
      journal = {\apjss},
         year = 2023,
        month = oct,
       volume = {268},
          eid = {55},
        pages = {55},
          doi = {10.3847/1538-4365/acef1b},
archivePrefix = {arXiv},
       eprint = {2308.06914},
 primaryClass = {astro-ph.SR},
       adsurl = {https://ui.adsabs.harvard.edu/abs/2023ApJS..268...55W}
}

@article{welsch2003magnetic,
  title={Magnetic helicity injection by horizontal flows in the quiet {S}un. {I}. {M}utual-helicity flux},
  author={Welsch, B T and Longcope, D W},
  journal={\apj},
  volume={588},
  pages={620},
  year={2003},
  publisher={IOP Publishing}
}

@ARTICLE{wiegelmann2021livrev,
       author = {{Wiegelmann}, Thomas and {Sakurai}, Takashi},
        title = "{Solar force-free magnetic fields}",
      journal = {\lrsp},
         year = 2021,
        month = dec,
       volume = {18},
          eid = {1},
        pages = {1},
          doi = {10.1007/s41116-020-00027-4},
archivePrefix = {arXiv},
       eprint = {1208.4693},
 primaryClass = {astro-ph.SR},
       adsurl = {https://ui.adsabs.harvard.edu/abs/2021LRSP...18....1W}
}

@article{zhang2010statistical,
  title={Statistical properties of solar active regions obtained from an automatic detection system and the computational biases},
  author={Zhang, Jie and Wang, Yuming and Liu, Yang},
  journal={\apj},
  volume={723},
  pages={1006},
  year={2010},
  publisher={IOP Publishing}
}

@article{zharkov2005statistical,
  title={Statistical properties of sunspots in 1996--2004: I. {D}etection, {N}orth--{S}outh asymmetry and area distribution},
  author={Zharkov, S and Zharkova, V V and Ipson, S S},
  journal={\solphys},
  volume={228},
  pages={377--397},
  year={2005},
  publisher={Springer}
}

\end{document}